\documentclass[aps]{revtex4}
\usepackage{graphicx}

\begin{document}

\title{Spiral Galaxies - classical description of spiral arms and rotational velocity pattern - toy model}
\author{Bogdan {\L}obodzi\'nski}
\email[]{bogdan.lobodzinski@desy.de}
\affiliation{Max Planck Institut F\"{u}r Physik, \\
Werner Heisenerg Institut, F\"{o}hringer Ring 6, 80805 M\"{u}nchen}

\date{\today}

\begin{abstract}
We propose an explanation of features of spiral galaxies: spiral arms and observed flat rotation curves, without
the presence of an exotic form of matter.
The formalism is based on Boltzmanns transport equation for the collisional matter and
the very-low-velocity post-Newtonian approximation of the general relativity equations
expressed in the Maxwell-like form.
The Maxwell-like formulation provides the base for the explanation of
the above phenomena in the language of dynamically created gravitoelectromagnetic fields by the movement of mass streams
in the plane of the galaxy disc.
In the model we use radical simplifications expressed as neglect of the gravitational interaction between neighbors
and approximation of the incompressible mass flow.
In this frame we show that if the galaxy disc is fuelled constantly by collisional mass carriers,
then the amplification of the gravitomagnetic field can be large enough to create the rotational velocity pattern
and spiral arms.
In this framework the collisional part of the mass gas in the galaxy disc plane, i.e. molecules and atoms,
is responsible for the creation of the gravitomagnetic field.
According to the model the spiral pattern of arms is static and determined by a direction of the mass flow. 
The model reproduces qualitatively the observed spiral arms and reproduces well the shape of the rotational velocity pattern.
As an example of the usability of the proposed mechanism, we reproduce qualitatively the above features
for the IC 342 and NGC 4321 galaxies.
\end{abstract}


\maketitle

\section{MOTIVATION}

Observational features of the spiral galaxies such as presence of spiral arms, rotational and radial velocity
patterns and the mass distribution in the disc remain without satisfying physical explanation. Moreover, the radial dependence of the
rotational velocity and the spiral pattern in galaxies are treated in the literature as completely unrelated subjects.

In case of the rotational velocity profile, most of the models developed in the last years make either extensive use of the concept of
dark matter \cite{darkmatter} or ideas coming from alternative theories to General relativity: scalar-tensor theories of gravity,
non-symmetric gravitational theory or modified Newtonian dynamics (MOND) \cite{mond1}.
Current explanations of the observational behavior of the rotational velocity which do not use dark matter concepts 
exploit numerical computation of the mass distribution in the galaxy discs as a function of the rotational velocity profiles
taken from the observational data (for example \cite{classical1}) or vice-versa. In fact, such a methodology does not explain
the tangential velocity behavior. It suggests only the connection between these two phenomena.

Explanations of the spiral arm pattern are based mostly on the density wave theory
\cite{classicaldensitywave}. More about the concepts of galaxy evolution developed during the last years can be found in \cite{concepts}.
A disagreement arises if, for example, the lifetime of the spiral arms is considered. Usually, the wave theory is based on the cool collisionless
discs of stars. However, such a construction generates rather short-lived transient spiral arms \cite{armsimul}.

As a result we have complex and fully independent models explaining each observational feature:
the spiral structure of galaxies and the velocity patterns of galaxy disc components.

In this work we develop a toy model in which the spiral spatial pattern and the behavior of rotational velocity of spiral galaxies are a consequence 
of the same physical mechanism.
The model explains the nearly flat circular velocity pattern and the spiral structure of the galaxy disc
using a gravitomagnetic field resulting from the post-Newtonian (PN) approximation of Einstein's gravity equations.

\section{OVERVIEW}

In the very-low-velocity PN approximation the gravity contains a velocity-independent and a frame and a velocity-dependent force. The latter,
so called gravitomagnetic force,
is similar to the forces generated by the electric and magnetic fields in classical electromagnetism.
The gravitomagnetic field is a feature of rotating massive objects (\cite{framedrag,ring}) and mass flows \cite{PNfluid}.
Working in the very-low-velocity regime of the PN approximation, Einstein's equations in harmonic gauge  can be recast in the Maxwell-like
form using variables corresponding to the gravitoelectric and gravitomagnetic fields \cite{DSX,gel1}.

Traditionally, in the description of large mass systems such as a galaxy, in a barycentric frame, the gravitomagnetic part is neglected due to the fact
that the magnitude of this component is usually associated with a rotating galaxy core only.
The gravitomagnetic field generated by the rotating galaxy core is much too small to be able 
to govern the motion of masses in the galaxy disc (far from the rotating object $\sim \frac{J_{0}}{r^3}$, where $J_{0}$ 
is the angular momentum of the rotating object).
In such a case the very-low-velocity PN correction approximately agrees with the Newtonian description . As a consequence, 
such a treatment requires dark matter for explanation of the rotational velocity pattern.

In our approach we postulate that a description of a spiral galaxy requires 
\begin{itemize}
\item an open system, e.g. we allow for income and outcome of collisional masses from the galaxy environment, 
\item an influence of the gravitomagnetic field created by the inward and outward streams of the mass flow on 
the kinematics of the galaxy system.
\end{itemize}
The latter point is a consequence of a fact that a spacetime geometry is modified and determined by the mass-energy and the mass-energy
currents relative to other mass. In other words, it means that the gravitoelectromagnetic forces cannot be eliminated by the Lorentz transformations.
Therefore, with a given mass flow entering the galaxy system we can associate the gravitomagnetic vector potential $\vec{A}_{g}$ and a mass current 
$\vec{J}_{g}$ . Both the variables can be treated as new dynamical variables suitable for description of the dynamics of the galaxy matter. 
Using this generalization, the metric coefficients $g$ in harmonic gauge and in the SI unit system 
for a particle in the barycentric frame of the galaxy system far from the center, we write as \cite{DSX}
\begin{eqnarray}\label{metric1}
&&
g_{00}= -1 + 2 \frac{V}{c^2}, \nonumber \\
&&
g_{0i} = - 4 \frac{A_{i}}{c}, \nonumber \\
&&
g_{ij} = \delta_{ij} \left( 1 + 2 \frac{V}{c^2} \right)
\end{eqnarray}
where $V$ is a scalar gravitoelectric potential and $A_{i}$ is a vector gravitomagnetic potential.
We can describe the scalar potential as a sum $V = \Psi_{0} + \Psi_{g}$, where $\Psi_{0}$ is a gravitational potential 
of the galaxy center and $\Psi_{g}$ the gravitational potential seen by the incoming/outcoming mass created by the mass in the galaxy disc.

The vector potential $\vec{A}_{g}$ is determined by the amount of masses and their velocities inside the galaxy
and acts back on a mass current $\vec{J}_{g}$. 
Later on, instead of the vector potential $\vec{A}_{g}$ we will use the gravitomagnetic field $\vec{H}_{g}$.
The Maxwell-like equations used for the mass streams in the galaxy disc allow taking into account the dependence between
the mass currents $\vec{J}_{g}$ and the field $\vec{H}_{g}$ (the vector potential $\vec{A}_{g}$).

In our description we propose that the spiral form is a result of a flow of collisional masses between the galaxy environment and the
galaxy center. In such a frame the gravitomagnetic field can be amplified by the mass stream affecting the rotational component
of the velocity and the spiral shape of the stream. The shape of streaming masses can be obtained by the ratio of the tangential and the radial
mass currents. The tangential component is created by the gravitomagnetic field and the radial part by the law of mass conservation
(the continuity equation).

We now address the question: how can such a potentially large gravitomagnetic field be created in the spiral galaxy system?
In our model, we have two mechanisms leading to the amplification of the gravitomagnetic field:
\begin{enumerate}
\item the gravitoelectromagnetic Maxwell-like equations may lead 
to the divergence of the gravitomagnetic field created by the collisional masses entering the galaxy.
\item the streaming motion of masses. This process is based on the interaction of the existing gravitomagnetic field
with the field of the mass flow. 
\end{enumerate}
The both mechanisms exist on the level of the Maxwell-like equations. 

In the first case, the gravitomagnetic field is characterised by the negative gravitomagnetic diffusivity which implies that 
the gravitomagnetic field will increase instead of being damped.
During the evolution in time the gravitomagnetic field generated in the galaxy system, powered by an external mass
may reach large values over a large space scale before its further growth is halted by the violation of the linearisation conditions
of the post-Newtonian approximation of the general relativity theory.

In the second mechanism, the growth of the gravitomagnetic field can be described as an  
interaction of the self-gravitomagnetic field generated by the moving mass carriers with the seed gravitomagnetic field 
(for example, due to rotation of center of the galaxy). It can lead to creation of a net field acting back 
on the mass carriers in the galaxy disc plane.
Assuming that the gravitomagnetic field of the galaxy core is initially oriented along an axis perpendicular to the plane 
of the galaxy disc, the resulting gravitomagnetic
field will follow the same orientation. As a result, after long enough time, the mass movement inside the galaxy disc is governed by the central
gravitational potential of the
galaxy core and the vector gravitomagnetic potential corresponding to the net gravitomagnetic field in the galaxy disc plane.
Due to the orientation of the gravitomagnetic vector potential along the $\phi$ direction in cylindrical coordinates,
the central gravitational
field does not disturb dynamical effects induced by the vector potential.
In such a geometrical configuration the canonical angular momentum of the mass in the galaxy disc, proportional to
the vector gravitomagnetic potential,
becomes a constant of motion and reflects the behaviour of the net gravitomagnetic vector potential.  

We show that such a model can describe well the rotational 
velocity pattern and the optical arms of spiral galaxies.
This is demonstrated by performing quantitative fits of our results to the
rotational velocity profiles and to the shape of the optical spiral arms of galaxies: IC0342 and 
NGC 4321 (M 100).
Both fits are satisfactory and show that the proposed model can explain the basic physical mechanism responsible
for the creation and behaviour of spiral galaxies without the need of exotic phenomena.


\section{MODEL DESCRIPTION}

For the measured average velocity of masses in galaxies in the range  $km\mbox{ }s^{-1}$, the ratio $v/c\approx 10^{-3}$ is small enough to
work within the first Post-Newtonian (1PN) approximation.
Let us consider a galaxy model, consisting of a system of a rotating black hole in the center and mass carriers
moving around the center in a two-dimensional disc geometry. The disc has a constant thickness denoted by $h$. 
For the sake of brevity, we will ignore gravitational and radiative effects linked to
the gravitational interaction of stars with their neighbors and with the interstellar gas.
Also all relativistic effects are neglected. 
Masses around the center of the system we will consider
as a 2-component mass gas: collisional (which is formed from molecules and atoms) and collisionless (which contains heavy elements: stars).
By density of particles in the disc we consider the surface density ($\rho$) related to the three-dimensional particle density ($\rho_{3d}$) 
and the thickness of the disc ($h$) as 
\begin{equation}\label{surfacedensity}
\rho_{3d} = \frac{\rho}{h}
\end{equation}
We use the SI unit system. 

In our model we postulate that the gravitoelectromagnetic fields can be dynamically modified by the mass currents generated
in the galaxy system. Therefore, working with galactic (barycentric) coordinates $x^{\mu}$ ($\mu=\left\{0,1,2,3\right\}$), 
the gravitoelectromagnetic fields become unknown functions which should be solved.
Following the formulation introduced in \cite{gel1} and 
using $\vec{E}_{g}$ and $\vec{H}_{g}$ as symbols for the new dynamical fields,
we can write the Maxwell-like equations as 

\begin{eqnarray}\label{gravmaxwell1}
&&
\vec{\nabla}\vec{E}_{g} = - 4 \pi m \frac{\rho}{h}, \label{gravmaxwell1a}\\
&&
\vec{\nabla} \times \vec{E}_{g}  = -\dot{\vec{H}}_{g}, \label{gravmaxwell1b} \\
&&
\vec{\nabla} \vec{H}_{g} = 0, \label{gravmaxwell1c} \\
&&
\vec{\nabla} \times \vec{H}_{g} = -\frac{16\pi G}{h c^2} \vec{J}_{g}+\frac{4}{c^{2}}\dot{\vec{E}}_{g},\label{gravmaxwell1d}
\end{eqnarray}
where $c$ is the speed of light, $G$ the gravitational constant, $m$ is the mass of collisional particle and 
the $\vec{J}_{g}$ is the surface mass current 
expressed in the very-low-velocity approximation as 

\begin{equation}\label{masscurrent} 
\vec{J}_{g} = \rho m \vec{v},
\end{equation}
with $\vec{v}$ being a velocity of the mass current.
Both gravitoelectromagnetic fields can be expressed 
through the gravitational potentials: the Newtonian scalar potential $\Psi_{g}$ and 
the gravitomagnetic vector potential $\vec{A}_{g}$

\begin{eqnarray}\label{gravpotentials1}
&&
\vec{E}_{g} = \vec{\nabla} \Psi_{g} +\frac{4}{c^2}\dot{\vec{A}}_{g},  \nonumber\\
&&
\vec{H}_{g} = -4 \vec{\nabla} \times \vec{A}_{g},
\end{eqnarray}

In addition to the Maxwell-like equations in the 1PN (\ref{gravmaxwell1a}-\ref{gravmaxwell1d}) 
the geodesic equation for the test mass particle with 
velocity $\vec{v} = \frac{d\vec{x}}{dt}$ in the very-low-velocity approximation can be rewritten in the form of the Lorentz-force like form

\begin{equation}\label{lorentz1}
\frac{d\vec{v}}{dt} = \vec{E}_{g} + 2 \vec{v} \times \vec{H}_{g}
\end{equation}

In the model, the stellar system is considered as an open system with a central part (core) 
having a massive black hole, 
an outer region (cylindrical disc) and an external gaseous (hydrogen) reservoir .
The regions are distinguished on the basis of the mass motion inside them.
\begin{itemize}
\item
The central part, the core, is created by a very massive object (for example a black hole) and a dense gas of mass carriers. 
The gravitational field created by the very massive black hole dominates the motion of masses in its direct neighborhood, 
and is strongly influenced by complex n-body dynamics, which cannot be approximated by our simplified two-dimensional analysis. 
Let us denote the radius of the core by $r_{1}$ and assume that the black hole in the core rotates. 
The rotating mass 
$M_{0}$ with angular momentum $\vec{J_{0}}$ generates the  gravitomagnetic field $\vec{H}_{0}$ at the point $\vec{r}$ far from the center 
($r \geq  r_{1}$) where 
$\vec{H}_{0} = \frac{G}{2 c^{2}} \frac{1}{r^{3}} \left(
\vec{J}_{0} - 3 \left( \vec{J}_{0}\cdot \frac{\vec{r}}{r}\right) \cdot \frac{\vec{r}}{r}\right)$. 
For simplicity we will use the gravitomagnetic field at the equatorial plane, where $\vec{H}_{0}$ has only a non-zero $z$-component.
\begin{equation}\label{H0core}
H_{0} = \frac{G}{2 c^2} \frac{J_{0}}{r^3} \mbox{ for } r >= r_{1}
\end{equation}

\item
the outer part of the stellar system model is defined as a disc with an external ($r_{2}$) and internal ($r_{1}$) radii and is an area 
where the n-body gravitational interaction can be neglected and the movement of the mass carriers can be affected by the
$\phi$-component of the gravitomagnetic vector potential created by the field $\vec{H}_{g}$ aligned along the assumed $z$-axis
determined by the angular momentum of the rotating core of the galaxy. 
This area can be fuelled or emptied by a flow of collisional masses to or from the external gas reservoir.  
The disc is considered as a two-dimensional object with constant thickness $h$ much smaller then the external radii ($h<<r_{2}$) 
filled with a two-component gas of mass carriers: collisional and non-collisional. 
\begin{itemize}
\item non-collisional gas: large masses like stars for which the time between collisions is comparable or longer then the lifetime of the universe.
\item collisional gas: molecules and particles having masses of the order of the Hydrogen atom and the time between collisions 
is small in comparison with the age of the universe. 
\end{itemize}

\end{itemize}

The outer part of the galaxy system as defined above is the subject of our analysis.
The axial symmetry introduced by the gravitomagnetic field breaks the spherical symmetry of the central gravity field
and results in the creation of the current density ${\vec{J}}_{g}$ in the plane perpendicular to the gravitomagnetic field.
For simplicity of our calculations we neglect any movement along the $z$-axis, limit our analysis only to the plane
$\left(r,\phi\right)$ in the cylindrical coordinate system and replace the $z$ derivative by the disc thickness. 

In this area, i.e. far from the galaxy center where the velocities of the colliding masses are non-relativistic, 
in addition to the set of eqs (\ref{gravmaxwell1}) we introduce in this region the gravitational analog of Ohm's law for 
the collisional gas 

\begin{equation}\label{lorentz0}
\vec{J_{g}} = \sigma \vec{E_{g}} + \mu \left( \vec{J_{g}} \times \vec{H_{g}}\right) .
\end{equation}
Using notation characteristic for electromagnetic phenomena,
$\sigma = m \rho \tau $ corresponds to the electromagnetic conductivity and $\mu = 2 \tau$ to the mass mobility.
The $\tau$ is the time between binary collisions of the mass carriers with mass $m$ 
averaged over thickness of the disc (the surface relaxation time). 
The eq. (\ref{lorentz0}) is derived from Boltzmann's transport equation in relaxational time approximation,
\begin{equation}\label{boltzmann1}
\partial_{t} f + \vec{v}\cdot \vec{\nabla} f + \left( - m\left(\vec{\nabla} \Psi_{0}\right) + \vec{F}_{g}\right) \cdot \partial_{\vec{p}} f = 
\left. \frac{df}{dt}\right|_{collisional},
\end{equation}
with the kinetic momentum $\vec{p}$ and the kinetic energy of the body with mass $m$ 
contributing to the flow expressed as $\epsilon = \frac{p^2}{2 m}$.
The force $\vec{F}_{g}$ is generated by the net gravitoelectric and the net gravitomagnetic fields:

\begin{equation}\label{lorentz2}
\vec{F}_{g} = m \vec{E}_{g} + 2 m \left( \vec{v} \times \vec{H}_{g}\right).
\end{equation}

The relaxation time approximation is equivalent to the replacement of the right side of the Boltzmann equation, eq ({\ref{boltzmann1}}), 
by $\frac{df}{dt}|_{collisional} \rightarrow - \frac{f-f_{0}}{\tau}$ where $\tau$ is the relaxation time constant of the perturbed state function $f$
with respect to its equilibrium state.
Using this substitution and the perturbative expansion of the deviation of the state function 
$f_{1}\left(\vec{r},\vec{p}\right)$ from its equilibrium $f_{0}\left(\vec{r},\vec{p}\right)$ 
\begin{equation}\label{expansion1}
f = f_{0}\left(\vec{r},\vec{p}\right) + f_{1}\left(\vec{r},\vec{p}\right),
\end{equation}
we get
\begin{equation}\label{boltzmann2}
\overbrace{
\left[
\partial_{t} f_{0} + \vec{v}\cdot \vec{\nabla} f_{0} - m\left(\vec{\nabla} \Psi_{0}\right) \partial_{\vec{p}} f_{0}
\right]}^{\mathrm{A1}} +
\overbrace{
\left[
\partial_{t} f_{1} + \vec{v}\cdot \vec{\nabla} f_{1}  - m\left(\vec{\nabla} \Psi_{0}\right) \partial_{\vec{p}} f_{1} +
\vec{F}_{g} \partial_{\vec{p}} f_{0} + \vec{F}_{g} \partial_{\vec{p}} f_{1} + \frac{f_{1}}{\tau}
\right]}^{\mathrm{A2}} = 0.
\end{equation}

We specify the function $f_{1}$ in more detail. Let us analyse the case without gravitoelectromagnetic components
when $\vec{H}_{g}=\vec{E}_{g} = 0$. 
Taking into account our assumption that the motion of the mass carriers is governed only by the 
central part of the galaxy system (the potential $\Psi_{0}$), 
so that it does not depend on the gravitational interactions with neighbors, we can write $f_{1} = 0$. 
As a proof of this let us set $\vec{F}_{g}=0$. 
In this case we have to consider only direct elastic collisions between masses. 
The collision changes the velocity of the colliding masses but not the potential $\Psi_{0}$ 
in which the movement takes place and which is constant. 
After collision, the system can be seen as an equilibrium state with another set of initial conditions ($\vec{r}$, $\vec{p}$).
Using the relaxation time approximation, the part $A2$ in eq. (\ref{boltzmann2}) can be solved independently of the part  $A1$.
The equation for $f_{1}$ is

\begin{equation}\label{boltzmann3}
\partial_{t} f_{1} + \vec{v}\cdot \vec{\nabla} f_{1}  - m\left(\vec{\nabla} \Psi_{0}\right) \partial_{\vec{p}} f_{1} + \frac{f_{1}}{\tau} = 0, 
\end{equation} 
and its solution can be written as

\begin{equation}\label{boltzmann4}
f_{1}\left( \vec{r},\vec{p},t \right) = e^{-\frac{t}{\tau}} f_{0}\left(\vec{r},\vec{p},t\right),
\end{equation}
where the function $f_{0}\left(\vec{r},\vec{p}\right)$ is the solution of the equation denoted as $A1$ in eq. (\ref{boltzmann2}).
For larger times $t >> \tau $ the function $f_{1} \rightarrow 0 $. 
It should be noted that the simplified model presented above is not true for the case in which the gravitational interaction with neighbors is 
non-negligible. In this situation each collision modifies not only the momentum but the effective gravitational 
potential as well. Such a modified potential generates a further change in the state function causing the system exhibit gravitational clustering
where signs of cooperative behaviour can be seen.

It is clear that the assumption $\vec{F}_{g} = 0$ determines also the Newtonian limit in our model. 
As was discussed above, the state of the galactic system is governed by part $A1$ of eq. (\ref{boltzmann2}). 
Since it represents the equilibrium state of the system, the mass current generated by such a state function is equal to $0$. 
Therefore, in the Newtonian limit 
of this model the gravitational analog of Ohm's law is:

\begin{equation}\label{lorentz01}
\vec{J_{g}} = 0.
\end{equation}

In the model, the gravitoelectromagnetic components introduce correlations, gravitoelectromagnetic filamentation, 
into the system which leads to a nonzero displacement 
of the distribution function $f\left(\vec{r},\vec{p},t\right)$ from its equilibrium state. 
The gravitoelectromagnetic filamentation
generates the nonzero mass current $\vec{J}_{g}$ proportional to the displacement of the state function $f_{1}\left(\vec{r},\vec{p},t\right)$.
The displacement function $f_{1}$ 
is limited to terms no higher than first order in all the gravitoelectric fields existing in the system.  
Therefore, using the following form for the deviation of the state function:  

\begin{equation}\label{boltzmann4a}
f_{1}\left(\vec{r},\vec{p}\right) = \tau \left( \partial_{\epsilon} f_{0}\right) \left( \vec{v}\cdot \vec{\Lambda}\right),
\end{equation}
where the unknown vector $\vec{\Lambda}\left(\vec{r}\right)$ corresponds to the gravitoelectrical field
and is created when the system is displaced from the equilibrium state, we can get  the form of the gravitational analog of 
Ohm's law, eq. (\ref{lorentz0}).
Detailed derivation of equation (\ref{lorentz0}) is presented in Appendix \ref{appendixBoltzmann}.

For the collisionless part of the galaxy disc components (stars) the gravitational Ohm's law is an improper approximation 
because of the very large relaxation time (nearly equal to the age of the universe). In this case, 
neglecting gravitational interaction with neighbors, the heavy masses in the system have to be considered as a 
subsystem in the equilibrium state.     
The perturbative expansion of the state function, defined in eq. (\ref{expansion1}) with correction $f_{1}$ given by 
eq. (\ref{boltzmann4a}) is no longer a valid approach.  
Therefore the collisionless gas of masses cannot create gravitoelectromagnetic fields, however due to the nonzero value of the mass, this fraction 
of the gas can interact with the gravitoelectromagnetic fields created by the collisional component of the galaxy system.

The equation of motion for each component of the galaxy system (collisional or non-collisional) in the outer part of the disc is governed by the 
same Lorentz-like formula, 

\begin{equation}\label{lorentz3}
\vec{F} = -m \left(\vec{\nabla} \Psi_{0}\right) + m \vec{E}_{g} + 2 m \left( \vec{v} \times \vec{H}_{g}\right),
\end{equation}
where the mass $m$ and the velocity $\vec{v}$ can be associated with a colliding particle or a star.  
The gravitoelectromagnetic parts $\vec{E}_{g}$ and $\vec{H}_{g}$
are the fields created by the collisional component of the galaxy disc described by the gravitational Ohm's law (\ref{lorentz0}).

Let us stress the difference between the Ohm-like law (\ref{lorentz0}) and its electromagnetic analog. Due to the fact that the mass $m$
plays the role of the charge, the conductivity in the system is proportional to the mass,
while the term proportional to the gravitomagnetic field is mass independent.
As a consequence of this all participating masses show
the same dynamics due to the gravitomagnetic fields. Equation (\ref{lorentz0}) is the fundamental one for the model presented here.

The creation and the initial development of the current stream described by equation (\ref{lorentz0})
requires more detailed considerations. For the purpose of this work we 
can list two possible mechanisms for the initialisation of the gravitomagnetic field related to the mass flow: 
\begin{itemize}
\item by the existence of the gravitomagnetic field generated by the massive rotating black hole in the center of the galaxy $H_{0}$ 
(eq. (\ref{H0core}),
\item by the gravitational n-body interaction in the galaxy bulge resulting in creation of the spatial stream pattern.
The pattern remains very unstable if we consider only the gravitational multi-body interaction \cite{galaxyevolution}.
\end{itemize} 
When the resulting gravitomagnetic field $\vec{H}_{g}$ is large enough, 
the field stands to act as a supporting mechanism for development of mass stream spatial structures.
Since the gravitomagnetic fields are weak, the above mechanism can be considered on small distances  
between the initialisation sources. The increase of the gravitomagnetic field allows to influence more distant masses.
Therefore, one should see an evolution of the galactic disc starting from a spatially small structure which 
grows with time. 

Another important assumption introduced in the model is incompressibility of the fluid where we  
treat the density of masses in the developed paths of generated currents as constant.
From now on, we will model a mass filament as a two-dimensional flow of an incompressible fluid and a nondissipational structure. 
In this limit the final form of the Maxwell-like equations describing the gravitoelectromagnetic fields associated with 
the mass current can be simplified to:

\begin{eqnarray}
&&
\vec{\nabla}\vec{E}_{g} = - 4\pi G m \frac{\rho}{h}, \label{gravmaxwell1finala}\\
&&
\vec{\nabla} \times \vec{E}_{g}  = -\dot{\vec{H}}_{g}, \label{gravmaxwell1finalb} \\
&&
\vec{\nabla} \vec{H}_{g} = 0, \label{gravmaxwell1finalc} \\
&&
\vec{\nabla} \times \vec{H}_{g} = -\frac{16\pi G}{h c^2} \vec{J}_{g},\label{gravmaxwell1finald} \\
&&
\vec{E}_{g} = - \vec{\nabla} \Psi_{g}, \label{gravmaxwell1finale} \\
&&
\vec{H}_{g} = -4 \vec{\nabla} \times \vec{A}_{g} . \label{gravmaxwell1finalf}
\end{eqnarray}

It is obvious that the gravitomagnetic field generated only by the rotating core of the galaxy is too weak to drive the mass carriers
in the outer part. However, additional mass flow - created spontaneously, can amplify the magnitude of the total 
gravitomagnetic field generated in the system. 
To see this let us take into account equation (\ref{gravmaxwell1finald}), in which the right part is proportional to the mass current. 
If we consider a loop of mass flow in the presence of an initial gravitomagnetic field $\vec{H}_{0}$ 
the gravitomagnetic field generated by the mass movement has the same direction as the initial field $\vec{H}_{0}$. 
This increase of the resulting gravitomagnetic field $\vec{H}_{g}$ leads to increase of the mass current.
As a result the system becomes diverging and in the limit violates the linearisation conditions. 
Working within the non-relativistic and post-Newtonian approximations and using
eqs (\ref{gravmaxwell1b}), (\ref{gravmaxwell1d}) and (\ref{lorentz0}) 
we derive the equation for the gravitomagnetic field $\vec{H}_{g}$ alone
(for details of derivation see appendix \ref{appendixHalone}),

\begin{eqnarray}\label{Halone1}
&&
\partial_{t}\vec{H}_{g}
- \frac{2}{\rho m} \left( \vec{\nabla} \times \left( \vec{J}_{g} \times \vec{H}_{g} \right) \right) =
-\Lambda \left( \vec{\nabla}^2 \vec{H}_{g} \right) \nonumber \\
&&
\mbox{ where  } \Lambda = \frac{h c^2}{16 \pi G} \frac{1}{\rho m \tau} . 
\end{eqnarray}

The positive coefficient $\Lambda$ with the present minus sign on the right side of the equation above has 
the character of a negative gravitomagnetic diffusivity.
The existence of the surface particle density $\rho$ and the surface relaxation time $\tau$ in the gas implies that the diffusivity coefficient $\Lambda$
shows a local character specific for each galaxy.
The negative diffusivity amplifies small perturbations (in our case it could be an initial gravitomagnetic field $\vec{H}_{0}$)  
and can lead to an explosion. It is a situation in which we are not restricted by
any limits. The calculations presented here are done in the frame of the very-low-velocity 1PN approximation and are valid only inside such limits. 

Summarizing, the basic idea of the model can be presented as follows.
Spontaneously created surface mass current flow starts to induce a gravitomagnetic self-field
which, interacting with the central gravitomagnetic field (or with the field created by another surface mass flow),
is magnified and leads to the gravitomagnetic net field which begins to modify the surface current.
An increasing gravitomagnetic field increases the mass involved in the flow. Because we assumed, that the net surface density of the current
remains constant, the length of the stream has to be increased, which causes an increase of the gravitomagnetic field.
The system starts to behave in a cooperative way, increasing the length of surface mass streams and grows spatially with time.

In the case of the Newtonian limit (eq. (\ref{lorentz01})), i.e. without incoming and outcoming matter from the gas reservoir, the
gravitomagnetic field in the system is due solely to the rotating black hole in the galaxy core ($\vec{H}_{0}$. eq. (\ref{H0core})).
The small value of the gravitomagnetic field $H_{0}$ allows us to completely ignore it for 
$r > r_{1}$ ($\vec{H}_{g} = \vec{H}_{0} \approx 0$).
Discussion of the resulting gravitomagnetic field based on the solution of eq. (\ref{Halone1}) and the possible 
mechanisms leading to slowing down the growth of the field are presented in the next section.


\section{RESULTING GRAVITOMAGNETIC FIELD IN THE GALAXY SYSTEM}
The gravitomagnetic field can be found by solving the induction equation (\ref{Halone1}). As a first step let us determine the 
radial component of the current ${J_{g}}_{r}$.  
Calculating the divergence from equation (\ref{gravmaxwell1d}) and using eqs (\ref{gravmaxwell1a}) 
we get the continuity equation 

\begin{equation}\label{continuity1}
-\frac{16\pi G}{h c^2} \vec{\nabla} \cdot  \vec{J}_{g}+16\pi \left( \partial_{t} \rho \right) = 0.
\end{equation}
Because we assumed that the mass stream acts as an incompressible fluid, the density of masses $\rho$ in the current is constant.
Therefore we can neglect any dependence of the density $\rho$ on the time or space variables. 
As a result we have the continuity equation in the form 

\begin{equation}\label{continuity}
\vec{\nabla} \cdot \vec{J_{g}} = 0 .
\end{equation} 
From this equation in cylindrical coordinates we can derive general solutions for the current ${J_{g}}_{r}$ 

\begin{equation}\label{eqjr}
{J_{g\epsilon}}_{r} = \epsilon \frac{1}{4\pi r} I_{0}\left(z\right),
\end{equation}
where $\epsilon=\pm 1$ and $I_{0}\left(z\right)$ is a general function describing a total mass current.
By limiting our analysis to the disc
plane, i.e. by neglecting any motion along the $z$-axis we will use $I_{0}$ as a constant variable, independent of the distance $r$ 
and the tangential angle $\phi$.  
The current $I_{0}$ is the external mass flow introduced into the galaxy system. 

Using equations (\ref{gravmaxwell1d}) 
we can determine directions of the current flow corresponding to the $\pm$ sign: 

\begin{itemize}
\item
${J_{g+}}_{r} = \frac{1}{4\pi r} I_{0}$ for the inward direction,
\item
${J_{g-}}_{r} = -\frac{1}{4\pi r} I_{0}$ for the outward direction.
\end{itemize}

These two possibilities suggest the existence of two spiral flow patterns (as combinations of the radial and tangential
components of the current). Each of them can be determined by the boundary conditions.
Each flow creates its own self gravitomagnetic field, which can interfere with the existing field. 
Such a resulting gravitomagnetic field affects back the mass flows.
In the case of the initialization of the 
filamentation process, the existing field is treated as an initial gravitomagnetic field generated by the rotating galaxy core. 
Additional modifications of the spiral shape are possible if the mass density inside the filament changes due to some external reasons.
We will not consider such situations.

Introducing the radial components of the mass current into the induction equation (\ref{Halone1}) and approximating the $r$-component of the 
gravitomagnetic field as ${H_{g}}_{r} = 0$, which was justified by the earlier assumption of neglecting the mass motion along the $z$-axis,
we get the differential equation for the $z$-component of the gravitomagnetic field ${H_{g}}_{z}$

\begin{equation}\label{Halone01zcomp}
\frac{\alpha \sigma}{h}\left(\partial_{t} {H_{g\epsilon}}_{z}\right) +
\left( \partial_{r}^{2} {H_{g\epsilon}}_{z} \right)  + \frac{1+\epsilon B_{0}}{r} \left( \partial_{r} {H_{g\epsilon}}_{z} \right) = 0,
\end{equation}
where the coefficients are

\begin{eqnarray}\label{betacoeffs}
&&
\epsilon=\pm 1,\mbox{  }\alpha = \frac{16\pi G }{c^{2}},\mbox{   the surface conductivity } \sigma = \rho m \tau,\nonumber \\
&&
\mbox{ and the cooperativity parameter } B_{0} = \frac{\alpha \mu I_{0}}{4 \pi h} = \frac{4 G \mu I_{0}}{h c^2} .
\end{eqnarray}

The time between collisions $\tau$, averaged over column length $h$ and aligned along $z$-coordinate could be estimated
by comparison with the time between collisions in the three-dimensional configurational space $\tau_{3d}$.
Approximating the collisional molecular gas by a classical hard-sphere gas one can define the time between elastic collisions as
\begin{equation}\label{coll3d}
\tau_{3d} \sim \frac{1}{\rho_{3d} v_{coll} \sigma_{coll}}
\end{equation}
where the index $3d$ denotes the three-dimensional space, $v_{coll}$ is a random 
velocity of the colliding hard-sphere and $\sigma_{coll}$($=\pi R^2$) is a cross-section for the collision of the hard-spheres with 
a radius $R$.

By analogy with the time $\tau_{3d}$ we can write a similar equation for the surface collisional time $\tau$ as
\begin{equation}\label{coll2d}
\tau \sim \frac{1}{\frac{\rho}{h} v_{coll} \sigma_{coll}}
\end{equation}
where we introduced the surface particle density $\rho$ and the thickness of the disc $h$. We assume the same random collisional velocity and the same
cross-section for elastic collisions. 
The above equations lead to the formula
\begin{eqnarray}
&&
\tau \sim h \frac{\rho_{3d}}{\rho} \tau_{3d} = \frac{h}{v_{h}},\label{coll2versus3d}\\
&&
v_{h}=\frac{\rho}{\rho_{3d}}\frac{1}{\tau_{3d}},\label{velcolumn}\\
\end{eqnarray}
where $v_{h}$ could be interpreted as an averaged velocity of a particle between succcesive collisions along the column, parallel to 
the $z$-axis and of a length $h$.

Using the variable $v_{h}$ we can eliminate the disc thickness $h$ from the equation (\ref{Halone01zcomp}), what 
results in a form
\begin{equation}\label{Halone02zcomp}
\alpha \sigma_{\Sigma}\left(\partial_{t} {H_{g\epsilon}}_{z}\right) +
\left( \partial_{r}^{2} {H_{g\epsilon}}_{z} \right)  + \frac{1+\epsilon B_{0}}{r} \left( \partial_{r} {H_{g\epsilon}}_{z} \right) = 0,
\end{equation}
where
\begin{eqnarray}
&&
\mbox{the surface conductivity } \sigma_{\Sigma} = \frac{\rho m }{v_{h}},\label{sigmacalc}\\
&&
\mbox{the cooperativity parameter } B_{0} = \frac{8 I_{0} G}{v_{h} c^2}\label{B0calc}.
\end{eqnarray}

The general solution of eq. (\ref{Halone01zcomp}) can be written as 

\begin{equation}\label{Halone01zsolution1}
{H_{g\epsilon}}_{z}\left(r,t\right) =  F\left(t\right) G_{\epsilon}\left(r\right),
\end{equation}
the factor $F\left(t\right)$ plays the role of the field amplitude and can be written in the form

\begin{equation}\label{Halone01amp}
F\left(t\right) = e^{\frac{\lambda^2}{\alpha \sigma_{\Sigma}} t}
\end{equation}
The argument $\lambda \left[\frac{1}{m}\right]$ is the integration constant and is a free parameter in our model.

The second ($r$-dependent ) factor is:
\begin{equation}\label{Halone01rdep}
G_{\epsilon}\left(r\right) = \left(\frac{r}{r_{eff}}\right)^{-\frac{B_{0}}{2}\epsilon} 
\left( C_{1} J_{-\frac{B_{0}}{2}\epsilon }\left( \lambda r \right) +
C_{2} J_{\frac{B_{0}}{2} \epsilon } \left( \lambda r \right) \right),
\end{equation}
where $C_{1,2}$ are the integration constants and $r_{eff}$ is the effective radius of the disc which will be determined 
together with the constants $C_{1,2}$.
The mass entering the galaxy system from the external gas reservoir 
appears in the gravitomagnetic field $\vec{H}_{g}$ (\ref{Halone01zsolution1}) in form of the cooperativity parameter $B_{0}$.

We see that, thanks to the contribution $F\left(t\right)$ (eq. (\ref{Halone01amp})), the field amplitude can become very large 
and strongly depends on the conductivity $\sigma_{\Sigma}$ of the filament.

Let us rewrite the general solution eq. (\ref{Halone01rdep}) in the form 

\begin{eqnarray}\label{Halone02rdep}
&&
G_{\epsilon}\left(r\right) = G_{0\epsilon} \left(\frac{r}{r_{eff}}\right)^{-\epsilon\frac{B_{0}}{2}} g\left(r \right) 
\mbox{  for  } \epsilon=\pm 1 \nonumber \\ 
&& 
\mbox{where  }
g\left(r \right) = J_{-\frac{B_{0}}{2}}\left( \lambda r \right) + J_{\frac{B_{0}}{2}}\left( \lambda r \right)
\end{eqnarray}
where we used $C_{1} = C_{2} = G_{0\epsilon}$ .

Due to the time dependence of the gravitomagnetic field we have to allow for discontinuity of the solution $G_{\pm}\left(r\right)$ at the core 
and the external radii ($r=r_{1,2}$). 
Therefore the parameters effective radius $r_{eff}$ and $G_{0\pm}$ have to be
determined from the conservation of the total gravitomagnetic flux crossing the disc which is limited by the outer and inner radii $r_{1}$ and
$r_{2}$ respectively.

In our case, for a given radius $r_{2}$ the flux conservation rule (with surface element $d \vec{s} = ds_{z} = r d\psi dr$) can be written as: 

\begin{equation}\label{Fluxconcervationlaw}
\int_{S} {\vec{H}}_{g} \cdot d\vec{s} = \int_{S} {\vec{H}}_{0} \cdot d\vec{s}.
\end{equation}
Using ${\vec{H}}_{g} = {\vec{H}}_{g+} + {\vec{H}}_{g-}$ as a total self gravitomagnetic field in the disc $r_{1} < r < r_{2}$ and
$\vec{J}_{0} = J_{0}\hat{z}$ we have:
\begin{equation}\label{Fluxconcervation1}
G_{0+} r_{eff}^{\frac{B_{0}}{2}} \int_{r_{1}}^{r_{2}} r f_{+}\left(r \right) dr + 
G_{0-} r_{eff}^{\frac{-B_{0}}{2}} \int_{r_{1}}^{r_{2}} r f_{-}\left(r \right) dr = 
H_{0}\left(r=r_{1}\right) {r_{1}}^3 \left( \frac{1}{r_{1}}-\frac{1}{r_{2}}\right)
\end{equation}
where 
\begin{eqnarray}\label{Fluxconcervation2}
&&
H_{0}\left(r=r_{1}\right)=\frac{G J_{0} }{ 2 {r_{1}}^3 c^{2} } \nonumber \\
&&
f_{\pm}\left(r \right) = F\left(t\right) r^{\mp\frac{B_{0}}{2}} g\left(r \right)
\end{eqnarray}
We impose on the field amplitudes $G_{0+}$ and $G_{0-}$ by the requirement that the gravitomagnetic flux
has the same value as the flux generated by the initial gravitomagnetic field $\frac{G J_{0}}{2 {r_{1}}^3 c^{2}}$ of the galaxy core.
The variable $J_{0} = \frac{G}{c} {M_{0}}^{2}$ is the maximum angular momentum of the rotating black hole with mass $M_{0}$. 
The mass of the black hole $M_{0}$ will be used as a parameter in numerical simulations of real galaxies.

The real solution for the variable $r_{eff}$ can be found for
\begin{equation}\label{constantscond1}
G_{0+} = - G_{0-} = \frac{G}{2 c^{2}} \frac{J_{0}}{{r_{1}}^{3}}.
\end{equation}
Under simplifying assumptions the solution for the effective radius $r_{eff}$ is:

\begin{equation}\label{effradius1}
r_{eff}^{\frac{B_{0}}{2}} = \frac{1}{2 I_{+}} \left( \eta+\sqrt{{\eta}^{2}+4 I_{+} I_{-}}\right)
\end{equation}
with
\begin{eqnarray}\label{wdefs}
&&
\eta=  \left( \frac{1}{r_{1}}-\frac{1}{r_{2}}\right){ r_{1}}^{3} \\
&&
I_{\pm} =  \int_{r_{1}}^{r_{2}} r f_{\pm}\left(r \right) dr
\end{eqnarray}

Looking into the dependence of the gravitomagnetic fields ${\vec{H}}_{g\pm}$ on the distance $r$ we see that for $r < r_{eff}$ the system is
dominated by the field ${\vec{H}}_{g+}$ associated with the inward flow of mass carriers $J_{g+}$. For the $r > r_{eff}$ - by the field
${\vec{H}}_{g-}$ created by the opposite current flow $J_{g-}$ (outflow). 
At $r=r_{eff}$ the net gravitomagnetic field $\vec{H}\left(r=r_{eff}\right)$ is $0$.
It is important to note that we can get sufficiently large gravitomagnetic field only in case when both the inwarding and 
outwarding mass streams are taking into account. 
In the situation when only a single mass flow is considered, for a time smaller than the age of the universe, the gravitomagnetic field is not 
magnified enough to modify the mass dynamics in the system.   

\begin{figure}[hltb]
\centering
\includegraphics[width=0.45\textwidth]{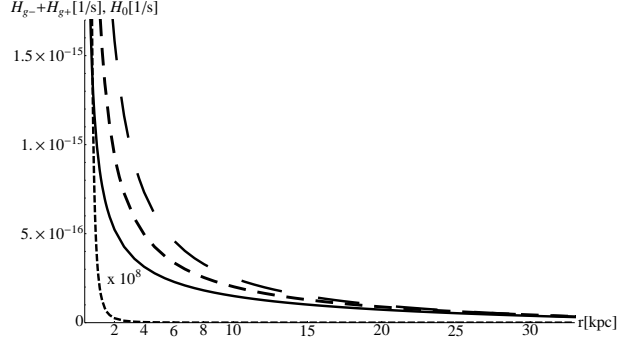} 
\caption{
Radial dependence of the gravitomagnetic field expressed as a sum of the gravitomagnetic fields generated by the inward and outward streams and
calculated for different cooperativity parameters:
$B_{0}=1.1$ (solid curve),
$B_{0}=0.9$ (short-dashed curve) and
$B_{0}=0.7$ (long-dashed curve).
For comparison we plot the field created by the galaxy core $H_{0}$, eq. (\ref{H0core}) which is multiplied by a factor $10^{8}$ for better
visualisation (dotted line).
The values of the parameters used for calculations are:
the evolution time  $t = 5\cdot 10^{16} \mbox{ }s \sim 1.58 \cdot 10^{9} \mbox{  }years$,
the center radius $r_{1}=1\cdot 10^{17} \mbox{ }m \sim 0.003 \mbox{ }kpc$, the external radius $r_{2}=100 \mbox{ }kpc$,
the parameter $\lambda=5.2\cdot 10^{-22} \mbox{ } m^{-1}$,
the mass of the galaxy center $M_{0}=2.6\cdot 10^{40}\mbox{ } kg$, the surface conductivity $\sigma_{\Sigma} = 42.62.24 \mbox{ }kg \mbox{ } s\mbox{ }m^{-2}$.
Effective radii $r_{eff}$ calculated for used $B_{0}$ are:
$48.86$, $51.2$ and $53 \mbox{ }kpc$ respectively.
\label{gravmagfield}}
\end{figure}

In Figure \ref{gravmagfield} we present results for total gravitomagnetic fields (${\vec{H}}_{g+}+{\vec{H}}_{g-}$) 
as a function of the distance $r_{1}<r<r_{eff}$ for 3 different values of the cooperativity parameter $B_{0}$. 
The full set of parameters used for the calculations is noted in the Figure caption.
For comparison, the gravitomagnetic field created by a rotating massive black-hole in the far field approximation (eq. (\ref{H0core}))
is of the order of $10^{-24}\mbox{ }s^{-1}$.
The dynamically generated gravitomagnetic field $H_{g}$ is 
approximately $10^{8}$ times larger than the gravitomagnetic core field $H_{0}$ .

Having a model of the creation of the gravitomagnetic field responsible for the spatial and tangential velocity patterns of the 
spiral galaxy model, we can ask for possible mechanisms which can stop the exponential growth of the field.
Precise analysis requires more detailed considerations. Here we describe a possible process qualitatively.

In agreement with our assumption that the galaxy system is open, 
increase of the gravitomagnetic field with time leads to increase of the galaxy disc potential generated by the incoming 
stream of masses from the external reservoir of the galaxy. This process violates the domination of the gravitational potential 
of the galaxy core $\Psi_{0}$ (the model presented here neglects the gravitational potential $\Psi_{g}$).
Such an increase of the gravitational potential of the system after a given period of time will modify the Lorentz-like equation of 
motion, eq. (\ref{lorentz1}), by introducing new terms in agreement with the full 1PN equation of motion.
The full geodesic equation for the test particle in 1PN approximation has the form \cite{gel1}
\begin{equation}\label{full1PN}
\frac{d}{dt}\left\{
\left[
1 + \frac{3}{c^2}\Psi_{g}+\frac{v^2}{c^2}
\right] \vec{v}
\right\}=
\left[
1-\frac{\Psi_{g}}{c^2}+\frac{3}{2}\frac{v^2}{c^2}
\right]\left( \vec{\nabla} \Psi_{g}\right)
+ \vec{v} \times \vec{H}_{g}.
\end{equation}

The gravitomagnetic filamentation process governed by the term $\vec{v} \times \vec{H}_{g}$ can be reduced by the nonlinear term 
proportional to $\Psi_{g} \left( \vec{\nabla} \Psi_{g}\right)$ in eq. (\ref{full1PN}):

\begin{equation}\label{nonlinear1PN}
\vec{E}_{g} = - \vec{\nabla} \Psi_{g} \rightarrow \vec{E}_{gNL} = \left( 1 - \frac{1}{c^2} \Psi_{g} \right) \left( \vec{\nabla} \Psi_{g}\right).
\end{equation}
Therefore, increase of the mass in the galaxy disc will result in 
a modification of the $r$-component gravitational Ohm's law eq. (\ref{lorentz0}),

\begin{equation}\label{Ohmmodification}
\vec{J_{g}} = \sigma \vec{E_{g}} + \mu \left( \vec{J_{g}} \times \vec{H_{g}}\right) \rightarrow 
\vec{J_{g}} = \sigma \vec{E}_{gNL},
\end{equation} 
which becomes free of growth instabilities.


\section{MOTION IN THE GRAVITOMAGNETIC FIELD}
\label{sec:Motion}

The resulting gravitomagnetic field ${H_{g}}_{z}={H_{g+}}_{z}+{H_{g-}}_{z}$
creates the $\phi$ component of the gravitomagnetic vector potential which is not dominated by the central gravitational
potential of the galaxy center.
If the considered mass system is governed only by the central gravity field,
then the system has spherical symmetry and is invariant under rotations around the center. In terms of
the average motion based on the state of the system, $f \left(\vec{r}, \vec{p}, t\right)$,  which fulfills
Boltzmann's transport equation (eq. (\ref{boltzmann1})), average values of radial and tangential components of stellar velocities are equal to 0.
An additional $z$-component of the gravitomagnetic field ${H_{g\mp}}_{z}$ breaks the symmetry, creating nonzero mean values of velocity components.
For the analysis of the velocity components of the colliding gas we use the Navier-Stokes momentum equation which, in the case of the assumptions of the model,
is equivalent to Jean's equations obtained from the Boltzmann transport equation (\ref{boltzmann1}).
Multiplying eq. (\ref{boltzmann1}) (written in cylindrical coordinates) by rotational component of the velocity, 
integrating over the velocity vector and averaging over the $z$-direction we get for the angular component of the fluid momentum
the following equation:

\begin{equation}\label{fluidmomentum1}
\rho \partial_{t} \left< v_{\phi} \right> +
\rho \left< v_{r} \right> \partial_{r} \left< v_{\phi} \right> + 2 \rho \left< v_{r} \right> {H_{g}}_{z} + 
\frac{\rho}{r} \left< v_{\phi} \right> \left< v_{r} \right> = 0,
\end{equation}
where $v_{\phi}$ and $v_{r}$ are the tangential and radial velocities of the flow respectively and $\rho$ denotes the surface
density of particles.
The collision term existing in the Boltzmann transport equation (eq. (\ref{boltzmann1})) averaged over velocities does not introduce
a non-zero term in the momentum equation. Assumed elastic collisions randomly distribute the velocities of colliding
particles.

In the 1PN approximation the radial mass current is expressed as $J_{gr\epsilon} = m \rho \left< v_{r\epsilon} \right>$.
The average radial velocity as the solution of the continuity equation (eq. (\ref{eqjr})) 
with help of eqs (\ref{masscurrent},\ref{eqjr} and \ref{B0calc}) can be rewritten to the form 
\begin{equation}\label{radialvel1}
\left< v_{r} \right>= \frac{\gamma}{r} \mbox{ where } \gamma = \frac{B_{0} c^{2}}{32 \pi G \sigma_{\Sigma} } .   
\end{equation}
Using this substitution, the equation for the rotational velocity of the mass fluid $\left< v_{\phi} \right>$ (eq. \ref{fluidmomentum1}) is 
\begin{equation}\label{vphi1}
\partial_{t} \left< v_{\phi} \right> +
\frac{\gamma}{r} \partial_{r} \left< v_{\phi} \right> +
\frac{\gamma}{r^2} \left< v_{\phi} \right> +
\frac{2 \gamma}{r} {H_{g}}_{z} = 0 .
\end{equation}
The solution of eq. (\ref{vphi1}) has the form
\begin{equation}\label{vphi2}
\left<\ v_{\phi} \right> = - 2 \frac{e^{-\epsilon \frac{\lambda^2}{B_{0}} r^2}}{r}
\int_{0}^{r} e^{\epsilon \frac{\lambda^2}{B_{0}}s^2}  {H_{g}}_{z} \left(s\right) s ds ,
\end{equation}
where we neglected the part proportional to $\frac{1}{r} e^{-\epsilon \frac{\lambda^2}{B_{0}} r^2}$ as it introduces not interesting behaviour.

Having both velocities, one obtains the trajectory equation
\begin{equation}\label{traj1}
\frac{d\phi}{dr} = 
-4 \left[ \epsilon \frac{16 \pi G}{c^2} \frac{\sigma_{\Sigma}}{B_{0}} \frac{1}{r} e^{-\epsilon \frac{\lambda^2}{B_{0}} r^2}
\int_{0}^{r} e^{\epsilon \frac{\lambda^2}{B_{0}}s^2}  {H_{g}}_{z} \left(s\right) s ds \right].
\end{equation}
We use this equation for the calculation of the trajectory of the test mass in the galaxy disc.
Equation (\ref{traj1}) shows that the mass carriers in a galaxy system do not rotate around the galaxy center on stable semicircular orbits.
Masses move along the spiral trajectories described by the differential equation (\ref{traj1}) determined
by the mass flow direction ($\epsilon$) and by the gravitomagnetic field resulting
from the cooperative behaviour of the mass carriers in the galaxy disc.

In the system geometry, the outward mass stream is characteristic for very large distances from the galaxy centre ($r>r_{eff}$).
Thus, we will concentrate our simulations on the space domain which could be comparable with observational data: 
the optical arms of galaxies and rotational velocity pattern. Both kind of data are determined for $r < r_{eff}$. 
Hence we will present our calculations for the inward mass current which is specific for this range of distance.  

\begin{figure}[hltb]
\centering
\begin{tabular}{cc}
\includegraphics[width=0.55\textwidth]{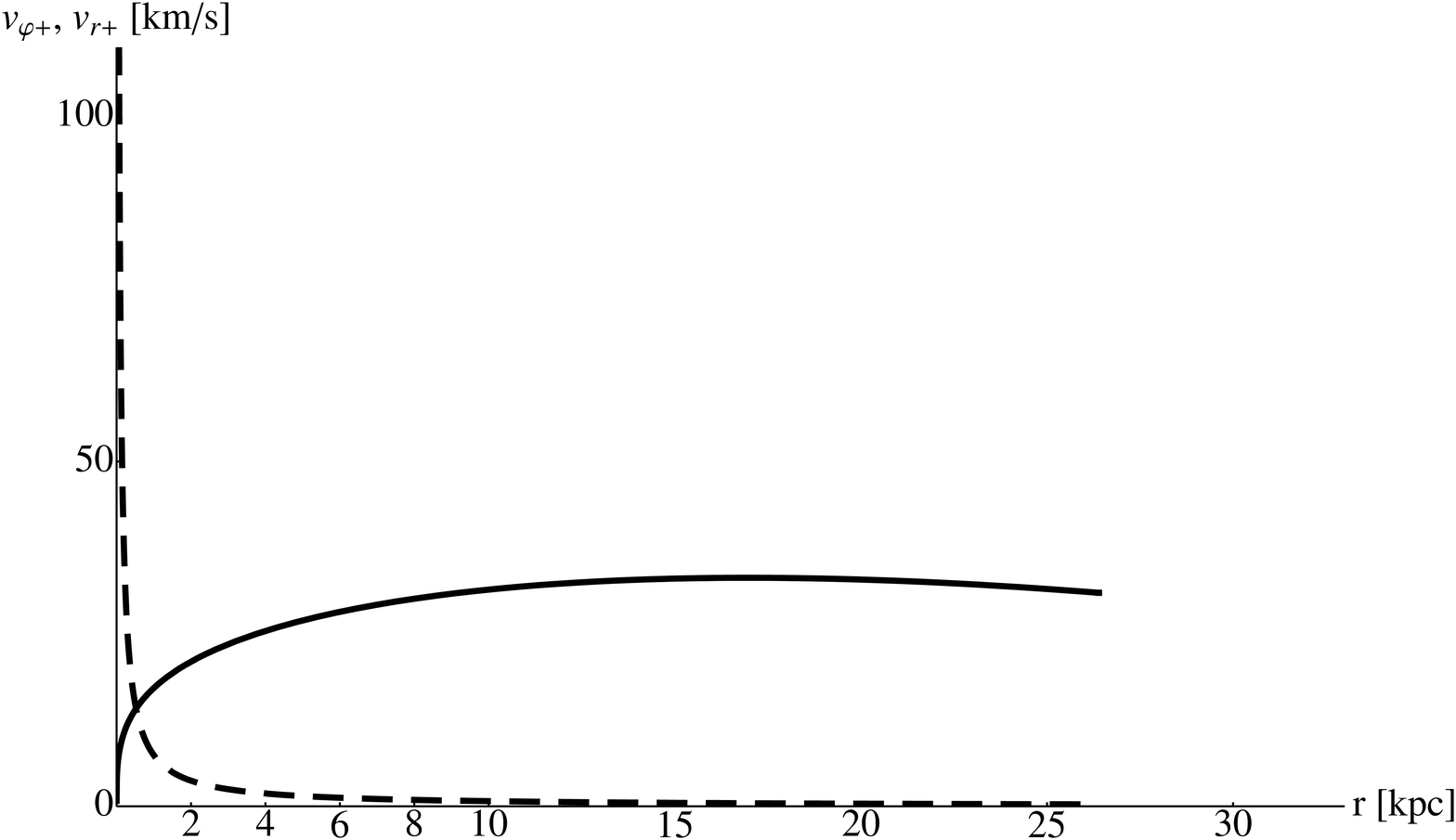} &
\includegraphics[width=0.4\textwidth]{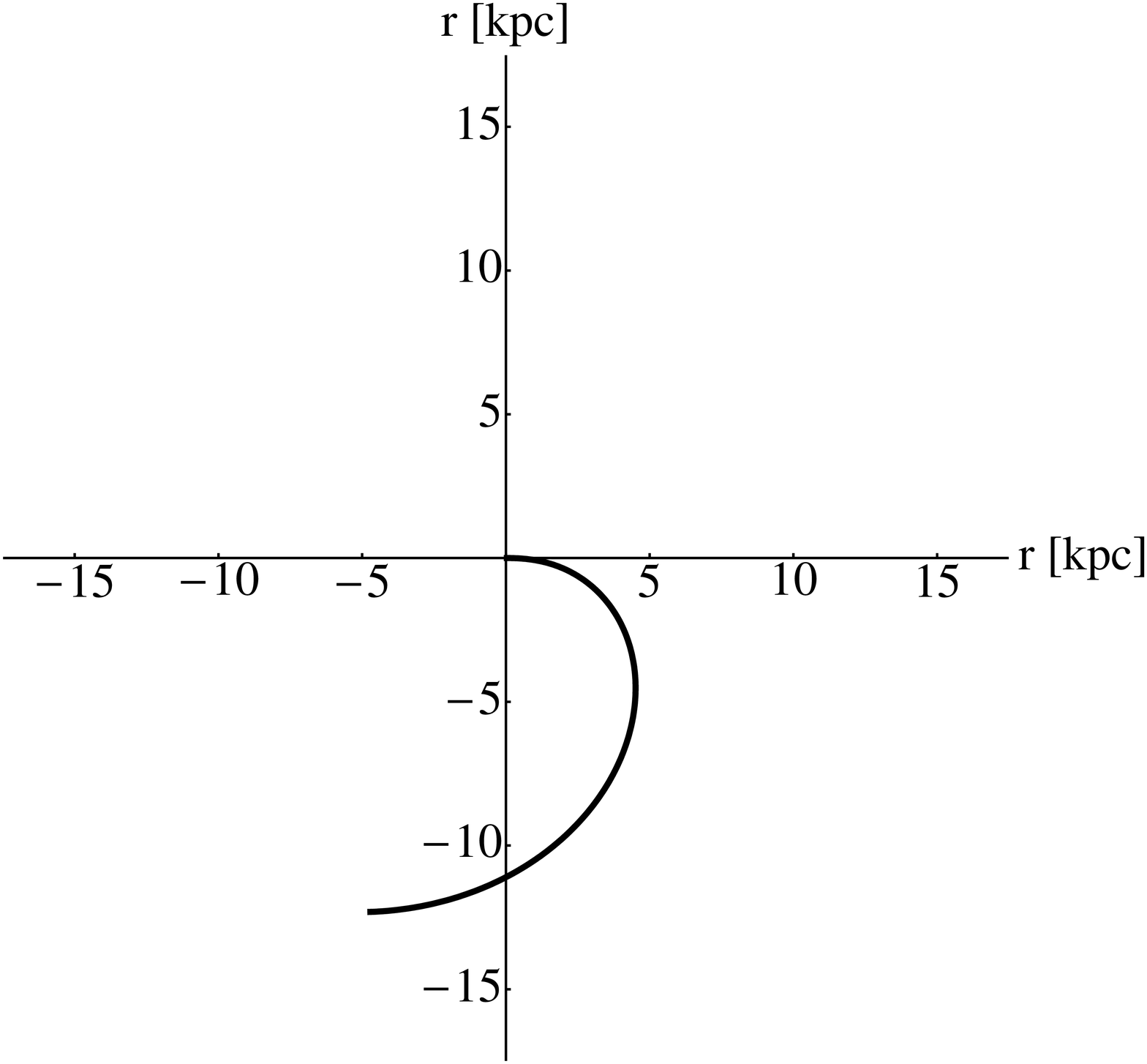}
\end{tabular}
\caption{
The velocity components and the trajectory of the inward mass flows as a function of the distance from the galaxy center
for the cooperativity parameter $B_{0}=0.7$.
Left panel: the tangential and radial profiles of velocities. The rotational velocity for inward flow $v_{\phi+}$
is denoted as a solid curve and the radial velocity ($v_{r+}$) as the broken curve.
Right panel: the trajectory of the test mass calculated using eq. (\ref{traj1})
corresponding to the inward streaming of masses.
All remaining values of parameters used for the simulation are the same as in Figure \ref{gravmagfield}.
\label{trajvis1}
}
\end{figure}

\begin{figure}[hltb]
\centering
\begin{tabular}{cc}
\includegraphics[width=0.55\textwidth]{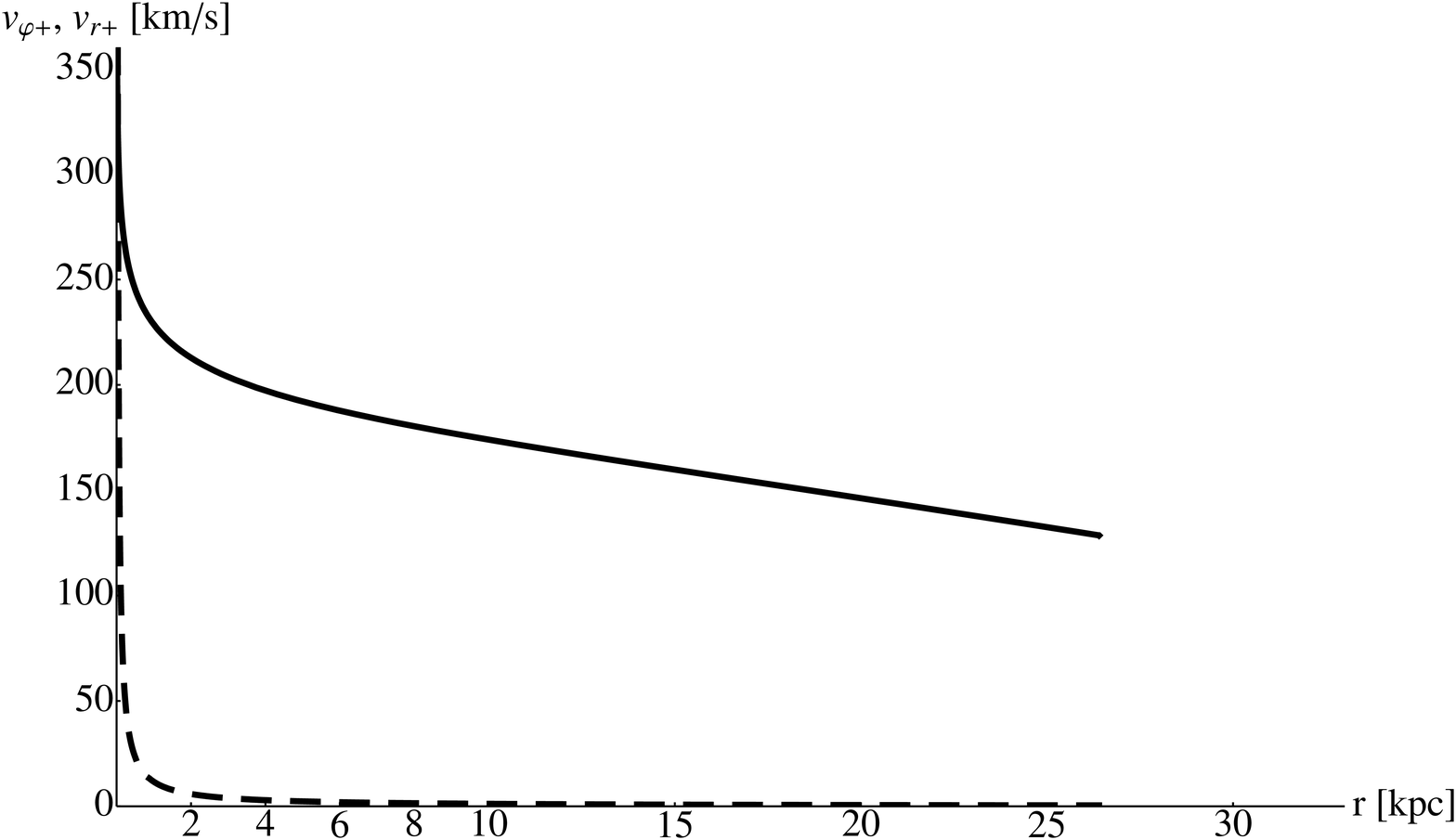} &
\includegraphics[width=0.4\textwidth]{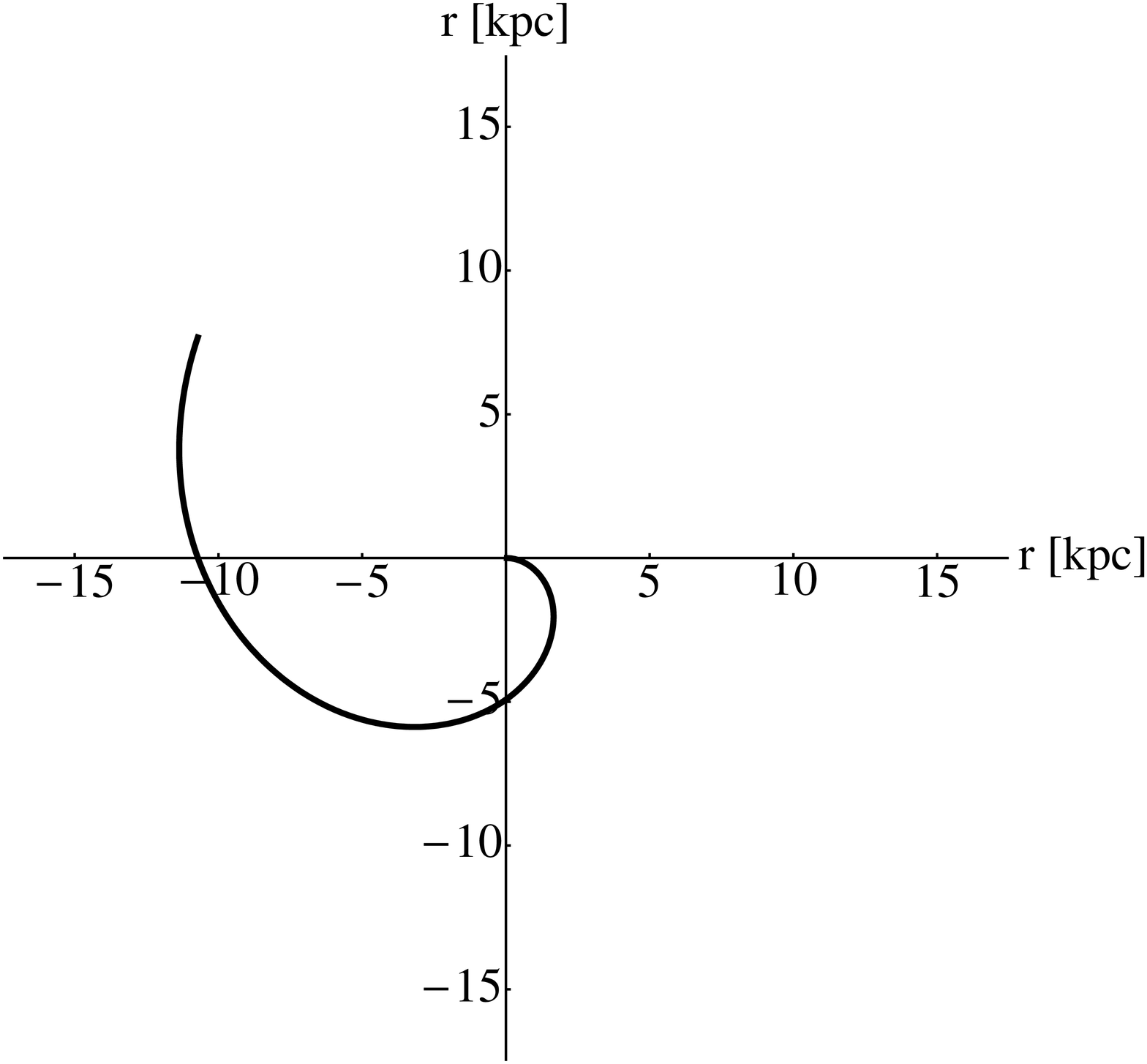}
\end{tabular}
\caption{
The same variables are presented as in Figure \ref{trajvis1} but with the cooperativity parameter
$B_{0} = 1.1$. All other values are the same as in Figure \ref{gravmagfield}. A diverging behaviour of the tangential velocity component
$v_{\psi+}$ is seen near the galaxy center. \label{trajvis2}
}
\end{figure}

In Figures  \ref{trajvis1} and \ref{trajvis2} we show the dependencies of the velocity components
of the inward stream  (rotational ${v_{\Phi}}_{+}$ and radial ${v_{r}}_{+}$) as a function of the distance
of the test mass from the galaxy center as well as the trajectory of the mass flow for the different
cooperativity coefficients $B_{0}$.
The mass is moving from periphery of the galaxy to the center .

The parameters $\lambda$ and the surface conductivity $\sigma_{\Sigma}$ are free parameters in the model. Therefore, the both parameters 
can be seen as internal features of the considered  galaxy. 
From the observational point of view, this means that the existence of galaxies with
very advanced spiral forms in the distant past is not ruled out.


\section{NUMERICAL RESULTS AND COMPARISON WITH OBSERVATIONS}
\label{sec:numcomparison}

The aim of this paper is to present a possible physical explanation of spiral galaxy features without additional complications 
due to non-analytic solutions. 
Therefore the model is highly idealized and therefore all comparisons of the theoretical results with existing observational 
data should be treated as qualitative only . 
In numerical calculations we assumed drastic simplifications, especially in the
neglect of the gravitational interaction between neighboring masses and by assuming an incompressible mass fluid.

The number of free parameters is rather large in our model. For the explanation of the spiral structure and 
the tangential velocity pattern we have to define the following constants:
\begin{itemize}
        \item 	the cooperativity parameter $B_{0}$, defined by eq. (\ref{betacoeffs}),
        \item 	the size of the inner and outer radii, $r_{1}$ and $r_{2}$ respectively,
	\item   the mass of the galaxy center $M_{0}$,
	\item   the collisional velocity $v_{h}$ (eq. \ref{velcolumn}),
        \item 	the scaling constant $\lambda$ responsible for the galaxy radius scaling and for the field amplitude, eq. (\ref{Halone01amp}),
	\item	the gravitational surface conductivity $\sigma_{\Sigma}$, eq. (\ref{sigmacalc}). For simplicity, we will use in our fits
		numerical values for $\sigma_{\Sigma}$ instead of considering separately the surface density $\rho$, the velocity $v_{h}$ and the mass $m$,
	\item   the evolution time of the galaxy $t$.
\end{itemize}

Due to the mapping of the disc evolution into the two-dimensional space we use the velocity $v_{h}$ (eq. \ref{velcolumn}) instead of 
the time between elastic collisions of particles. A value of the velocity can be estimated using a standard an observational data: the surface particle density 
$\rho \sim 10^{20} m^{-2}$, the average three-dimensional particle density $\rho_{3d} \sim 10^{12} m^{-3}$ (what corresponds to the average mass in the disc 
$\approx 10^{-14} \mbox{  } kg/m^{3}$) and the time between collisions $\tau_{3d} \sim 10^{14} \mbox{ } s$. For those values we have 
$v_{h} \sim 10^{-6} \mbox{ } m/s$. 
 
If the model is capable of simulating spiral galaxies, we should be able to reproduce 
the spiral pattern and the radial and rotational velocity pattern of a real galaxy with a given set of the above parameters.
For comparison of the predictions of our model with observed data we use data collected for the
IC0342 and NGC 4321 galaxies.
Both galaxies belong to the group Sc and have trailing arms. The observed data of the IC 342 and 
NGC 4321 galaxies are taken from \cite{rotationalcurves}.
The spatial trajectories of the simulated galaxies are corrected for inclinations given in \cite{rotationalcurves}.

\subsubsection{IC 342}
\label{subsec:IC342}

The galaxy IC0342 is an intermediate spiral galaxy classified as type Sc.
We can get an approximate agreement between calculated and observed tangential velocities and spiral
patterns for the following set of parameters:\\
the cooperativity coefficient $B_{0}=0.85$, the centre radius $r_{1}=0.003 \mbox{ }kpc $,
the external radius $r_{2} = 133.3\mbox{ }kpc$ and  
the mass of the galaxy core $M_{0} = 3.5 \cdot 10^{40} \mbox{ }kg$ ($= 1.75\cdot 10^{10}\mbox{ }M_{Sun}$). 
The surface conductivity is $\sigma_{\Sigma} = 94.7 \mbox{ } kg \mbox{ } s\mbox{ }m^{-3}$ what is achieved for 
the surface particle density $5.28\cdot 10^{21} \mbox{ }atoms\mbox{ }m^{-2}$ of hydrogen atoms ($2\cdot 10^{-27}\mbox{ }kg$) and the
velocity along the $z$-axis $v_{h} = 1.2\cdot 10^{-7}  \mbox{ } m s^{-1}$. 
The parameter $\lambda = 5.5\cdot 10^{-22} \mbox{ }m^{-1}$ and the time of the evolution $t = 5 \cdot 10^{16} \mbox{  }s\sim 1.58 \cdot 10^{9} \mbox{ }years$.
Calculated effective radius is: $r_{eff}=60 \mbox{ }kpc$.

\begin{figure}[hltb]
\centering
\begin{tabular}{cc}
\includegraphics[width=0.55\textwidth]{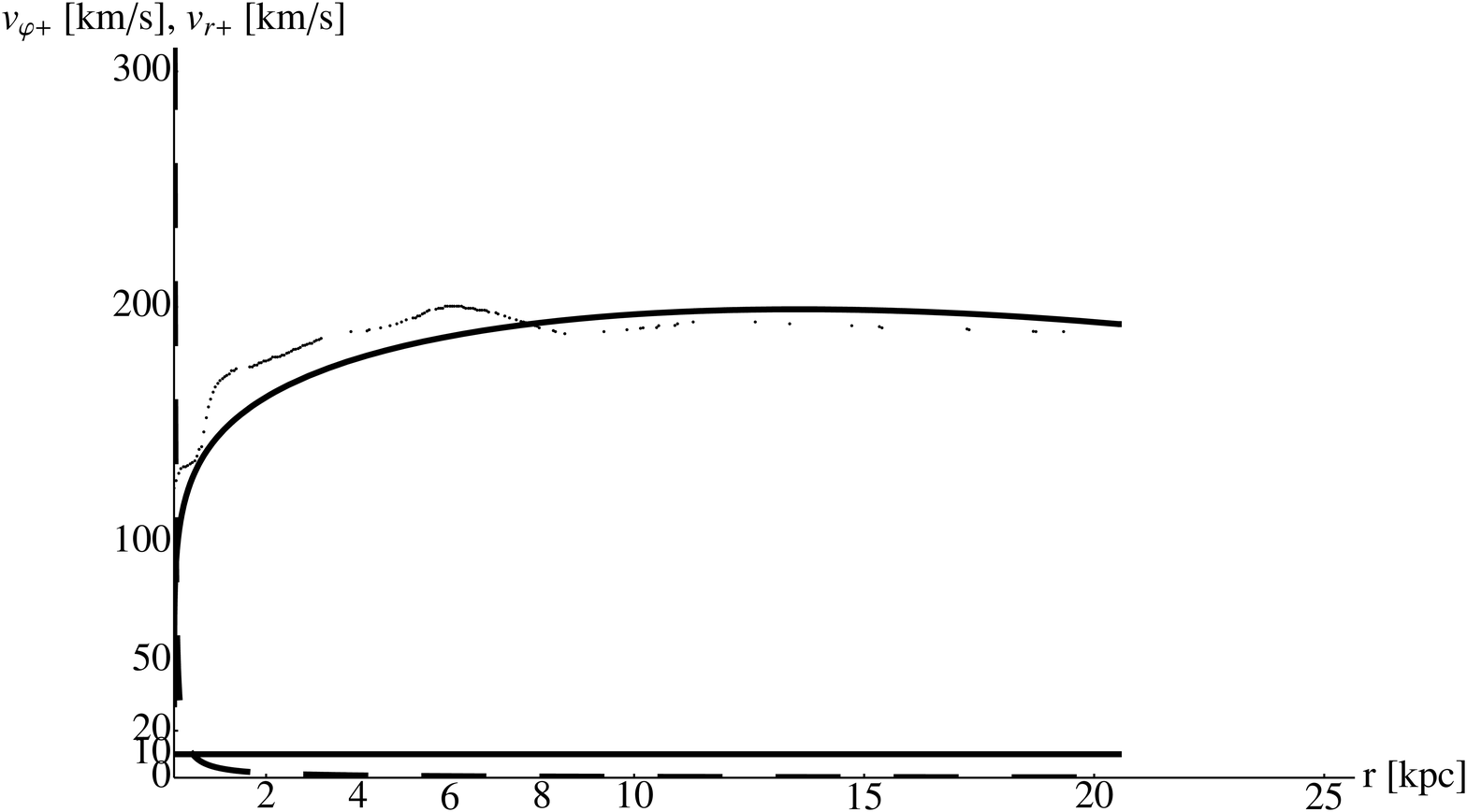} &
\includegraphics[width=0.4\textwidth]{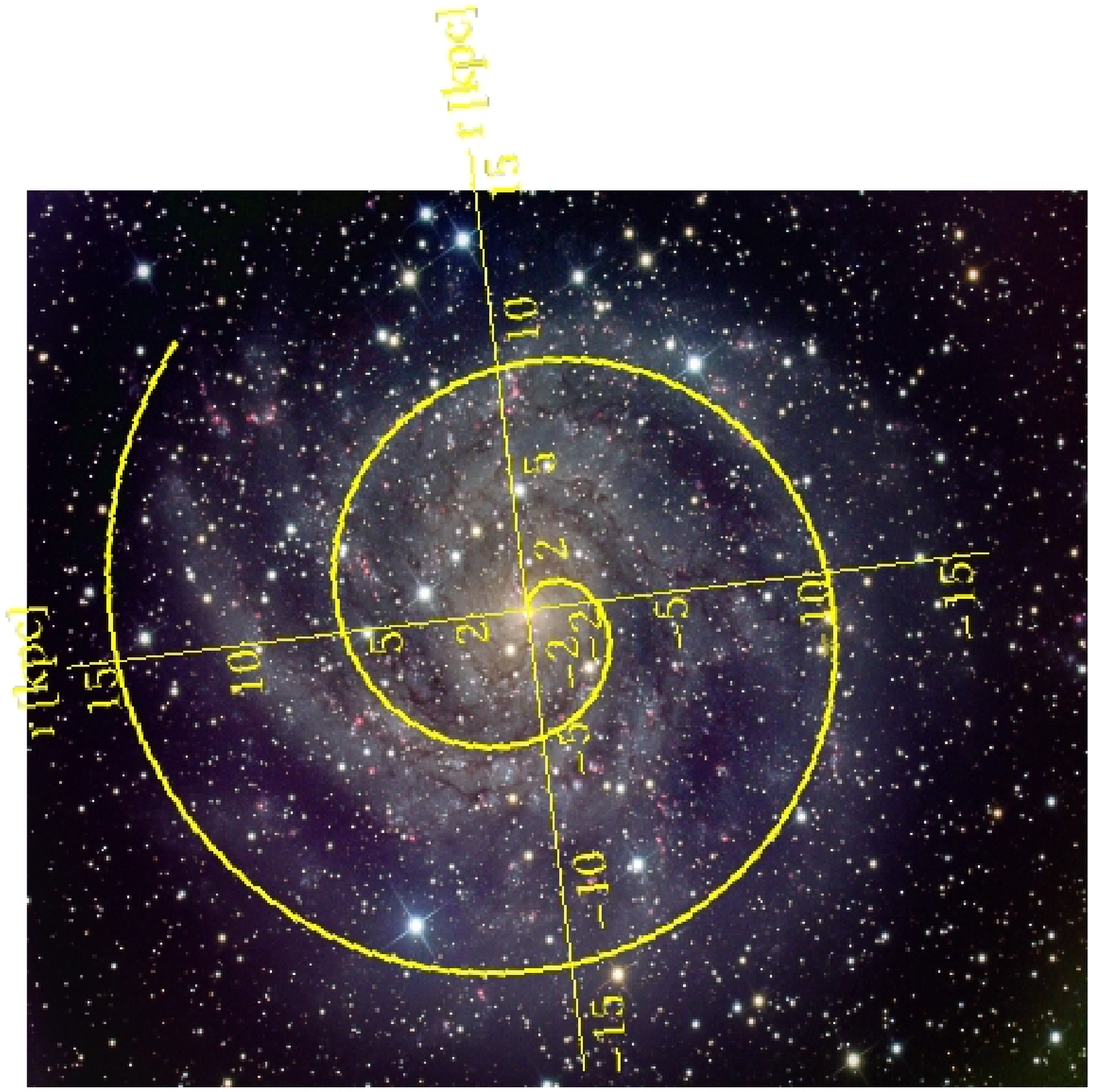}
\end{tabular}
\caption{
Left panel:
rotational (solid line) and radial (dashed line) components of velocity of a test mass for inward mass flow.
Observational data are given by black dots. The radial velocity at distances $r>1\mbox{ }kpc$ is smaller than
$10\mbox{ }km\mbox{ }s^{-1}$. The level $10\mbox{ }km\mbox{ }s^{-1}$ is denoted as a straight solid black line. 
Right panel:
comparison of the simulated spiral arm created by the inward mass stream with a photo of IC 342 
(Reproduced with permission from Bertrand Laville, http://www.deepsky-drawings.com/ic-342/dsdlang/en).
The values of all parameters used for the calculations are listed in the text.
\label{IC342comp}
}
\end{figure}

Our comparison is shown in Figure \ref{IC342comp} with the calculated and measured \cite{rotationalcurves}
inward flow velocity patterns and simulation of a single galaxy arm in the case of the inward mass current.

Due to the simplifications used in our model, the rotational patterns for small distances from the
inner radius cannot be recovered properly.
In the range of the bulge and dense halo radius ($< 3 \mbox{ } kpc$) the observed trajectory is not properly simulated. 
This may be explained by a large fraction of the elements near the galaxy center having a mass much heavier than that of hydrogen where 
the incompressible fluid approximation is not a proper assumption.
We can get approximate agreement for distances from the galaxy center larger than $3 \mbox{ }kpc$. 

The model does not take into account galaxy centers with a non-spherical shapes.   
In such cases more complicated dynamics are expected, which disturb significantly the movement close to the center area of the galaxy. 

\subsubsection{NGC 4321 (M 100)}
\label{subsec:M100}

The rotational velocity profile of the galaxy NGC 4321 is well documented in \cite{rotationalcurves}.
A similarity between simulated and observational data has been found for the following set of parameters:
the cooperativity coefficient $B_{0}=0.8$, the centre radius $r_{1}=0.003 \mbox{ }kpc $ ($1\cdot 10^{17} \mbox{ }m$), 
the external radius $r_{2}=117 \mbox{ }kpc$, the mass of the galaxy core
$M_{0} = 4.45\cdot 10^{40} \mbox{ }kg$ ($ 2.2\cdot 10^{10}\mbox{ }  M_{Sun}$), 
the surface conductivity  $\sigma_{\Sigma} = 42.62 \mbox{ } kg \mbox{ } s\mbox{ }m^{-3}$ (which corresponds to the surface particle density 
$2.6\cdot 10^{21} \mbox{ }atoms\mbox{ }m^{-2}$ of hydrogen atoms and the velocity $v_{h} = 1.2\cdot 10^{-7} \mbox{ } m \mbox{ }s^{-1}$), 
the parameter $\lambda = 5.7\cdot 10^{-22} \mbox{ }m^{-1}$ and the time of the evolution
$t = 5\cdot 10^{16} \mbox{ }s\sim 1.58 \cdot 10^{9} \mbox{ }years$). Calculated effective radius is $r_{eff}=56 \mbox{ }kpc$.

\begin{figure}[hltb]
\centering
\begin{tabular}{cc}
\includegraphics[width=0.55\textwidth]{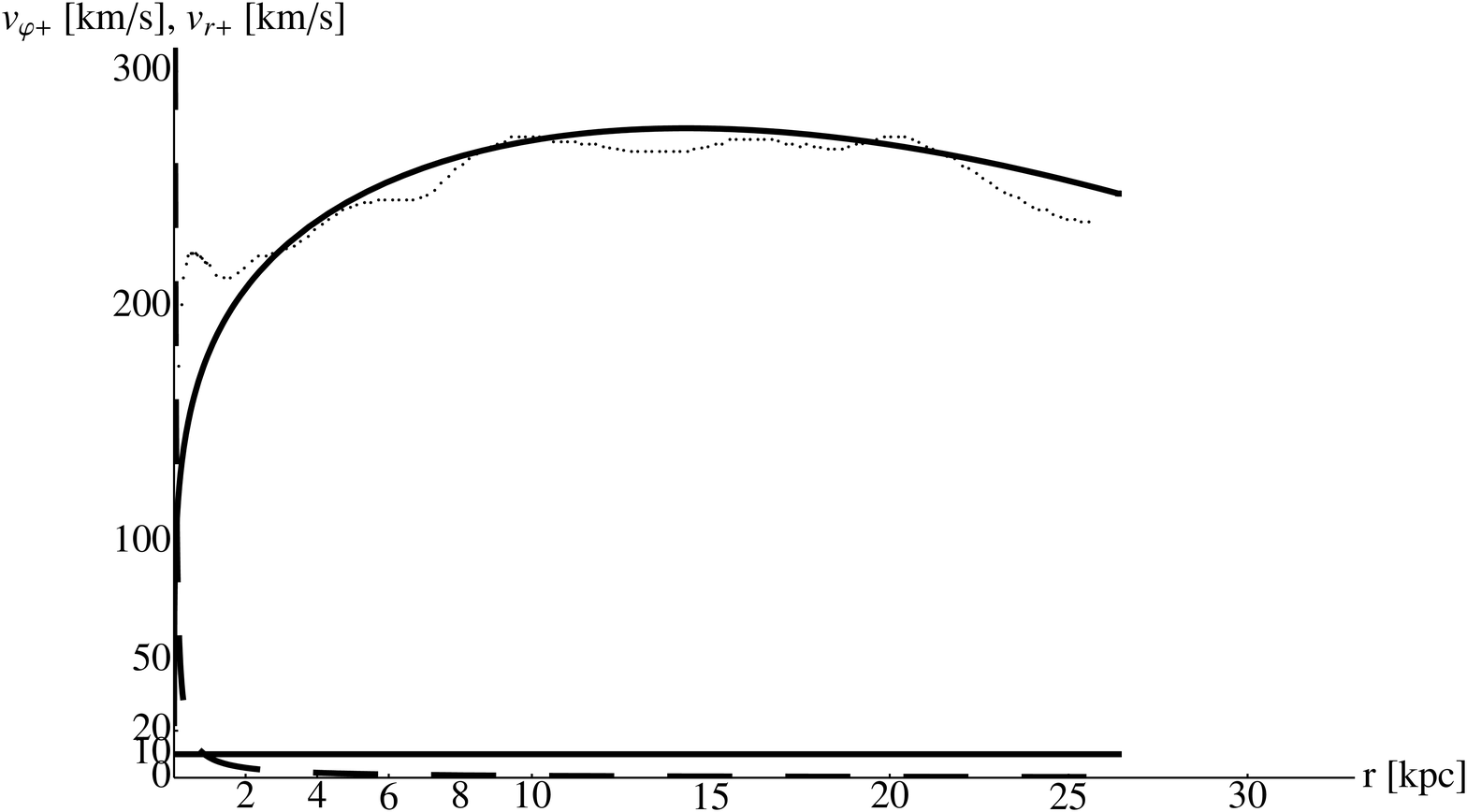} &
\includegraphics[width=0.4\textwidth]{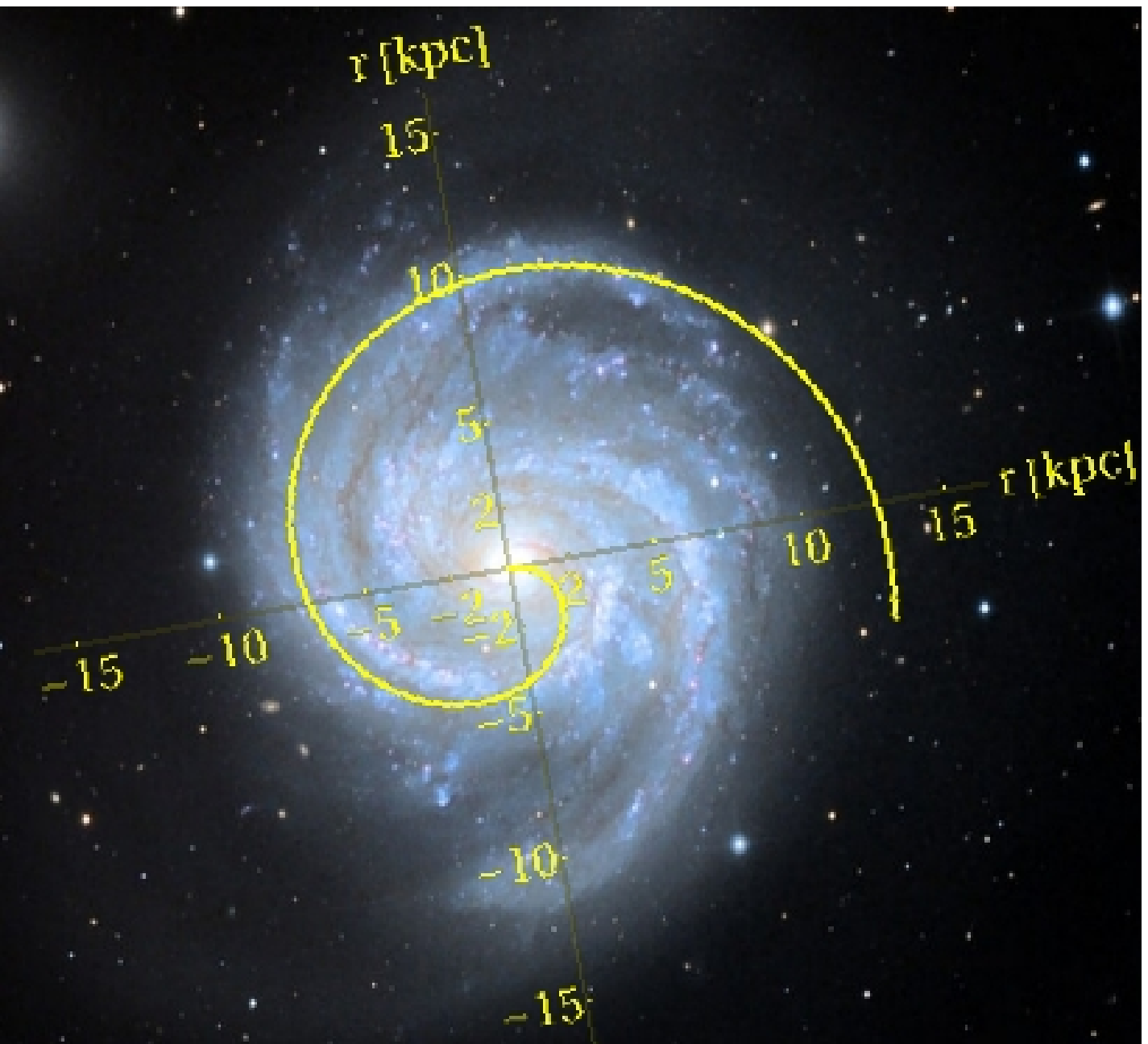}
\end{tabular}
\caption{
Left panel:
rotational (solid line) and radial (dashed line) components of velocity of a test mass for inward mass flow.
Observational data are given by the black dots.
The radial velocity reaches the value $< 10\mbox{ }km\mbox{ }s^{-1}$
for $r > 1\mbox{ }kpc$. The limit $10\mbox{ }km\mbox{ }s^{-1}$ is denoted as a straight solid black line.
Right panel:
comparison of the simulated spiral arm created by the inward mass stream with a photo of the M 100 galaxy
(Image courtesy of Joseph D. Schulman, http://en.wikipedia.org/wiki/Messier\_100)
The values of all parameters used for the calculations are listed in the text.
\label{NGC4321comp}
}
\end{figure}

The comparison of our simulations with the observational data is presented in Figure \ref{NGC4321comp},
which shows the calculated and the measured \cite{rotationalcurves}
rotational velocity patterns for the inward mass current and the simulation of a single galaxy arm.

As in the cases of the previous galaxy, IC0342, in the range of the center and the dense halo radius ($\sim 3 \mbox{ }kpc$)
our model does not provide a reasonable result. 
The comparison of the trajectory simulation with the observed arms shows an inconsistency in the area close to the galaxy center. 
We are able to simulate a nearly correct trajectory for $r>3\mbox{ }kpc $. 
As in the case of IC0342, we believe that this lack of agreement between simulation and observed data 
results from the dynamics of the galaxy center being more complex than assumed .


\section{CONCLUSIONS}

The galactic rotation curves and the mass distribution remain open questions to present day astrophysics. 
So far, these features have been explained by the introduction of dark matter
distributed accordingly in a spherical halo around the center of the galaxy. 
Existing alternative theories, which provide possible explanations of the noted galaxy features, like modified Newtonian dynamics (MOND) 
and generalized gravity \cite{genaralizedgrav}, predict a breakdown of Einstein's (and Newtonian) gravity at the scale of galaxies.  
A common problem for such theories is the lack of direct evidence of the required components (dark matter) and lack of other 
experimental verification. 

The galactic spiral pattern is usually treated as an independent phenomenon and is explained using the density wave theory 
\cite{classicaldensitywave}.

In the present work we show an alternative view, which gives a quantitative description of
the spiral pattern and the tangential and radial velocity behaviour of the spiral galaxies. 
The described model is only based on classical physics and uses the very-low-velocity first post-Newtonian approximation of general relativity,
which currently is also used successfully for a description of massive binary systems \cite{KNCT}.
Thus, we do not introduce any new kind of fields, exotic particles or other new phenomena. 
However, it should be stressed that the model does not remove the concept of dark matter. 
According to the model, all forms of matter interacting gravitationally (including dark matter) 
obey the presented evolution, leading to formation of the spiral structure and rotational velocity pattern.

The present model shows that the dynamics at the galaxy scale can be governed by the mass flow described by the Boltzmann's transport 
theory and by the velocity-dependent gravitational fields existing in the formulation of the post-Newtonian approximation 
of Einstein's gravity equation in the Maxwell-like form and therefore called gravitoelectromagnetic fields. 
Furthermore, the dynamics is governed by the cooperative behaviour of the participating collisional gas of mass carriers 
instead of pure Newtonian mechanics which acts as a $0$-th order interaction. 
The model introduces a physical mechanism responsible for both the spatial and rotational velocity patterns of spiral galaxies. 
The measured galactic rotation velocity and the existing spiral pattern of arms 
are a consequence of a well defined mechanism based on the cooperative behaviour of mass carriers. 
Mass flows, initialized for example by the gravitomagnetic field of the rotating galaxy core or 
by the spontaneous n-body gravitational interaction, start to develop into stable spiral mass streams 
due to the  self gravitoelectromagnetic fields. 
Due to the axial symmetry of the created vector potential of the self generated gravitomagnetic 
field, the rotational velocity pattern is a result of a classical constant of motion 
and depends on the cooperative parameter $B_{0}$, eq. (\ref{betacoeffs}).  
According to the author's best knowledge, there is no other work where the gravitoelectromagnetic cooperative 
behaviour of the mass carriers is taken into account for a description of the spiral galaxies.

Due to the drastic simplifications introduced into the model, making it a very preliminary description, 
definitions of some parameters could be difficult for unambiguous identification with observational data.

Basic simplifying assumptions are: 
\begin{itemize}
\item the galaxy is an open system,
\item the gravitational and radiative effects linked to
the gravitational interaction of stars with their neighbors and with inter-stellar gas and their possible relativistic effects are neglected,
\item the density of matter along the current paths is assumed constant (approximation of incompressible fluid),
\item the motion along the $z$-coordinate is neglected,
\item the value of the cooperativity coefficient $B_{0}$ is taken to be constant for the entire range of radius $r$.
\end{itemize}

The main predictions provided by the model are related to the existence of spiral galaxies in the 
universe and they can be summarized as follows:
\begin{itemize}
\item the galaxy exchanges the collisional mass with the external environment. 
\item the existence of collisional mass carriers is necessary for the creation of persistent spiral arms, 
\item the spiral arms of the galaxy are persistent structures. All masses in the disc area moves along the spiral trajectories, 
\item the radial velocity of the mass flow is below 10 $km\mbox{ }s^{-1}$, in agreement with observations,
\item the evolution of the galaxies is very individual. Thus implies that the spiral structure cannot be characterized by the 
 age of the universe, it is rather a specific feature of each galaxy. The existence of spiral galaxies 
in the early universe is not forbidden.
\end{itemize}

Despite all approximations, the presented model reproduces qualitatively 
such galaxy attributes as the shape of the spiral arms and the rotational 
velocity pattern as shown by comparison with observational data for the galaxies 
IC 342  and NGC 4321.



\section*{Appendices}

\subsection{}
\label{appendixBoltzmann}

In this appendix we derive the gravitational Ohm's law (eq. (\ref{lorentz0})).
We start with eq. (\ref{boltzmann1}) in the non-relativistic limit where the kinetic momentum $\vec{p}$ and 
the kinetic energy $\epsilon$ of the body with mass $m$ is expressed
as $\epsilon = \frac{p^2}{2 m}$  

\begin{equation}\label{appendixboltzmann1}
\partial_{t} f + \vec{v}\cdot \vec{\nabla} f + \left( -m \left(\vec{\nabla} \Psi_{0}\right) + \vec{F}_{g}\right) \cdot \partial_{\vec{p}} f = 
\frac{df}{dt},
\end{equation}
where the force $\vec{F}_{g}$ is (eq. (\ref{lorentz2})) 

\begin{equation}\label{appendixlorentz2}
\vec{F}_{g} = m \vec{E}_{g} + 2m \left( \vec{v} \times \vec{H}_{g}\right) . 
\end{equation}
Usage of the kinetic momentum is justified by the invariance of the state function $f\left(\vec{r},\vec{p},t\right)$ under exchange of the canonical and 
kinetic momentum and by the fact that the Lorentz-like force $\vec{F}_{g}$ is associated with the kinetic momentum.

Using the relaxation time approximation, the part $\frac{df}{dt}$ in the right side of eq. (\ref{appendixboltzmann1}),  
can be replaced by $\frac{df}{dt} \rightarrow - \frac{f-f_{0}}{\tau}$, what gives

\begin{equation}\label{appendixboltzmann12}
\partial_{t} f + \vec{v}\cdot \vec{\nabla} f + \left( -m \left(\vec{\nabla} \Psi_{0}\right) + \vec{F}_{g}\right) \cdot \partial_{\vec{p}} f =  
- \frac{f-f_{0}}{\tau} . 
\end{equation}

The gravitoelectromagnetic force $\vec{F}_{g}$ introduces the gravitoelectromagnetic correlation into the system (filamentation),
which perturbs the state function $f\left(\vec{r},\vec{p},t\right)$. 
In the linear approximation one can therefore write

\begin{equation}\label{appendixboltzmannsol1}
f = f_{0}\left(\vec{r},\vec{p}, t\right) + f_{1}\left(\vec{r},\vec{p}, t\right),
\end{equation}
where $|f_{1}| \ll |f_{0}|$.
The linearized Boltzmann transport equation (\ref{appendixboltzmann1}) for the equilibrium and 
for the perturbed distribution functions $f_{0}$ and $f_{1}$ becomes
\begin{equation}\label{appendixboltzmann13}
\overbrace{
\left[
\partial_{t} f_{0} + \vec{v}\cdot \vec{\nabla} f_{0} - m \left(\vec{\nabla} \Psi_{0}\right) \partial_{\vec{p}} f_{0}
\right]}^{\mathrm{A1}} + 
\overbrace{
\left[
\partial_{t} f_{1} + \vec{v}\cdot \vec{\nabla} f_{1}  - m \left(\vec{\nabla} \Psi_{0}\right) \partial_{\vec{p}} f_{1} +
\vec{F}_{g} \partial_{\vec{p}} f_{0} + 
\vec{F}_{g} \partial_{\vec{p}} f_{1} + \frac{f_{1}}{\tau}
\right]}^{\mathrm{A2}} = 0
\end{equation}
where we distinguish between the equilibrium component $A1$ and the perturbative term $A2$.

The gravitoelectromagnetic force $\vec{F}_{g}$ enters eq. (\ref{appendixboltzmann13}) together with the velocity 
gradient $\partial_{\vec{v}}$ of the $f_{1}$ component of the state function. 
Let us define the function $f_{1}$ as linear in the gravitoelectric fields and exact in terms of the gravitomagnetic field.
Because the potential of the galaxy center $\Psi_{0}$ is not modified by the galaxy disc potential and because the gravitoelectromagnetic
filamentation is created by the force $\vec{F}_{g}$, we can rewrite part $A2$ in eq. (\ref{appendixboltzmann13}) 
keeping only those terms which are directly proportional to the filamentation terms. In other words we consider 
the linearized Boltzmann transport equation (\ref{appendixboltzmann12}) for the perturbed state function $f_{1}$ taken along the unperturbed trajectories 
of masses in the central potential $\Psi_{0}$,
\begin{equation}\label{appendixboltzmann14}
A2 = \frac{d f_{1}}{dt} = 
\vec{F}_{g} \left( \partial_{\vec{p}} f_{0} \right) + 
\vec{F}_{g} \left( \partial_{\vec{p}} f_{1}\right) + 
\frac{f_{1}}{\tau}  = 0 .
\end{equation}

Using the formula 
\begin{equation}\label{appendixperturbation1}
\underline{\partial_{\vec{p}} f_{0}} = \left( \partial_{\vec{p}} \epsilon \right)  \left( \partial_{\epsilon} f_{0} \right) =
\left( \partial_{\vec{p}} \left( \frac{p^2}{2 m}\right) \right)   \left( \partial_{\epsilon} f_{0} \right) =
\underline{ \frac{\vec{p}}{m} \left( \partial_{\epsilon} f_{0} \right)},
\end{equation}
the vector identity $\left( \vec{v} \times \vec{H}_{g} \right) \cdot \vec{v} = 0 $
and the exact form of $\vec{F}_{g}$ (eq. (\ref{appendixlorentz2})), 
we obtain the perturbed part $A2$

\begin{equation}\label{appendixboltzmann15}
A2 = 
\left( \vec{E}_{g}\cdot \vec{p} \right) \left( \partial_{\epsilon} f_{0} \right) + 
\left(  m\vec{E}_{g} + 2 m \left( \vec{v} \times \vec{H}_{g} \right) \right)\cdot \left( \partial_{\vec{p}} f_{1}\right) +
\frac{f_{1}}{\tau}  = 0.
\end{equation}
Now, let us define
\begin{equation}\label{appendixboltzmannsol16}
f_{1}\left(\vec{r},\vec{p}\right) = \tau \left( \partial_{\epsilon} f_{0}\right) \left( \vec{p}\cdot \vec{\Lambda}\right)
\end{equation}
where the unknown vector $\vec{\Lambda}\left(\vec{r}\right)$ corresponds to the gravitoelectric field
created when the system is displaced from the equilibrium state. 

Replacing the velocity gradient by the energy derivative
\begin{equation}\label{appendixperturbation2}
\underline{\partial_{\vec{p}} f_{1} } = \partial_{\vec{p}} \left[ \tau \left( \partial_{\epsilon} f_{0} \right) 
\left( \vec{p} \cdot \vec{\Lambda} \right) \right] =
\tau \left[ \partial_{\vec{p}} \left( \partial_{\epsilon} f_{0} \right) \right]  \left( \vec{p} \cdot \vec{\Lambda} \right) +
\tau \left( \partial_{\epsilon} f_{0} \right) \vec{\Lambda} = 
\underline{ \tau \frac{\vec{p}}{m} \left( \frac{\partial^2 f_{0}}{\partial \epsilon^2} \right)  \left( \vec{p} \cdot \vec{\Lambda} \right) +
\tau \left( \partial_{\epsilon} f_{0} \right) \vec{\Lambda} }
\end{equation}
and introducing $f_{1}$ (eq. (\ref{appendixboltzmannsol16})) in eq. (\ref{appendixboltzmann15}), we have

\begin{eqnarray}\label{appendixboltzmann17}
&&
A2 =
\left( \vec{E}_{g}\cdot \vec{p} \right) \left( \partial_{\epsilon} f_{0} \right)
+ \overbrace{ \vec{E}_{g} \cdot \left( 
\tau m \vec{p} \left( \frac{\partial^2 f_{0}}{\partial \epsilon^2} \right)  
\left( \vec{p} \cdot \vec{\Lambda} \right)
\right) }^{neglected} \nonumber \\
&&
+ \overbrace{ m \vec{E}_{g} \tau \left( 
\left( \partial_{\epsilon} f_{0} \right) \vec{\Lambda}
\right) }^{neglected}
+ 
\overbrace{ \left( 
\vec{v} \times \vec{H}_{g} 
\right) \cdot 
\left( \tau  \frac{\vec{p}}{m} 
	\left( \frac{\partial^2 f_{0}}{\partial \epsilon^2} 
	\right)  
	\left( \vec{p} \cdot \vec{\Lambda} 
	\right) 
\right) }^{=0} \nonumber \\
&&
+ \left( 
\vec{v} \times \vec{H}_{g} 
\right) \cdot \tau 
\left( \partial_{\epsilon} f_{0} 
\right) \vec{\Lambda}  
 + \left( \partial_{\epsilon} f_{0} \right) \vec{p}\cdot\vec{\Lambda} = 0 .
\end{eqnarray}
Introducing the vector identity 
$\vec{\Lambda}\cdot \left(\vec{v} \times \vec{H}_{g}\right) = - \vec{v} \cdot\left( \vec{\Lambda} \times \vec{H}_{g}\right)$,
taking into account the overbraced actions and including $\vec{p} = m\vec{v}$ we get

\begin{equation}\label{appendixboltzmann18}
A2 =
\left[ \vec{E}_{g} + \vec{\Lambda} - \tau \left( \vec{\Lambda} \times \vec{H}_{g} \right) 
\right] \cdot \vec{v} \left( \partial_{\epsilon} f_{0} \right) = 0 .
\end{equation}
For each nonzero velocity $\vec{v}$ we have the condition 

\begin{equation}\label{appendixboltzmann19}
\vec{E}_{g} + \vec{\Lambda} - \tau \left( \vec{\Lambda} \times \vec{H}_{g} \right) = 0
\end{equation}

In Appendix (\ref{appendixCurrent}) we derive the expression for the surface mass density current 
\begin{equation}\label{appendixboltzmann22}
\vec{J}_{g} = - \sigma \vec{\Lambda},
\end{equation}
where the conductivity $\sigma$ is determined by the surface density of particles  $\rho \left(\vec{r}\right)$ as
\begin{equation}
\sigma  = \tau m \rho\left(\vec{r}\right) .
\end{equation}
Multiplying eq. (\ref{appendixboltzmann19}) by $\sigma$, we obtain the form (\ref{lorentz0})
\begin{equation}\label{appendixboltzmann23}
\vec{J_{g}} = \sigma \vec{E_{g}} + \mu \left( \vec{J_{g}} \times \vec{H_{g}}\right)
\end{equation}
with $\mu = 2 \tau$, completing our derivation.

\subsection{}
\label{appendixCurrent}

Here we derive the expression for the mass density current appearing in the paper as equation (\ref{appendixboltzmann22}).
\begin{equation}\label{appendixCurrent1}
\vec{J}_{g} = - \sigma \vec{\Lambda}
\end{equation}
where $\sigma = m \rho \tau$ is the surface mass conductivity. 
 
In terms of the energy-momentum tensor, the mass flux (current) is
\begin{equation}\label{appendixCurrent2}
\vec{J}_{g} = \frac{T^{0i} \hat{e}_{i}}{c}
\end{equation}
In the 1PN approximation and two-dimensional space we have 

\begin{equation}\label{appendixCurrent3}
\frac{T^{0i} \hat{e}_{i}}{c} = \int p^{i} \hat{e}_{i} f \left(\vec{p},{\vec{r}}\right) d^{2} p .
\end{equation}
Using the form $f = f_{0} + f_{1}$, where $f_{1} = \tau \left( \vec{p}\cdot \vec{\Lambda}\right) \left( \partial_{\epsilon} f_{0}\right)$,
we have

\begin{equation}\label{appendixCurrent4}
\frac{T^{0i} \hat{e}_{i}}{c} = 
\int p^{i} \hat{e}_{i} f_{0} + \tau \int p^{i} \hat{e}_{i}  
\left( \vec{p}\cdot \vec{\Lambda}\right) \left( \partial_{\epsilon} f_{0}\right) d^{2} p,  
\end{equation}
where the first part on the right hand side is equal to $0$ since it describes stationary conditions.

From appendix \ref{appendixBoltzmann}, we have 
$\frac{\vec{p}}{m} \left( \partial_{\epsilon} f_{0} \right) =  \partial_{\vec{p}} f_{0}$ (eq. (\ref{appendixperturbation1})), 
where $\vec{p} = p^{i} \hat{e}_{i}$, therefore eq. (\ref{appendixCurrent4}) becomes

\begin{equation}\label{appendixCurrent5}
\frac{T^{0i} \hat{e}_{i}}{c} = \tau m \int  \left( \vec{p}\cdot \vec{\Lambda}\right) \left( \partial_{\vec{p}} f_{0}\right) d^{2} p .
\end{equation}

We now use the general formula for scalar functions $f$ and $g$

\begin{equation}\label{appendixCurrent6}
\int_{V} f \left( \nabla_{p} g \right) dV = \int_{S} \left( g f \right) \hat{n} dS - \int_{V} g \left(\nabla_{p} f\right)
\end{equation}
where $V$ and $S$ stand for the volume and the surface enclosing the volume respectively.
With the condition 

\begin{equation}\label{appendixCurrent6boundary}
\int_{S} \left( g f \right) \hat{n} dS \rightarrow 0 
\end{equation}
and $\vec{T}^{0} = T^{0i} \hat{e}_{i}$,
the form (\ref{appendixCurrent5}) can be rewritten as

\begin{equation}\label{appendixCurrent7}
\frac{\vec{T}^{0}}{c} = -\tau m \int f_{0} \left( \nabla_{\vec{p}} \left( \vec{p}\cdot \vec{\Lambda} \right) \right) d^{2} p .
\end{equation}
Therefore, if we assume that the field $\vec{\Lambda}$ does not depend on the velocity $\vec{v}$ we have

\begin{equation}\label{appendixCurrent8}
\frac{\vec{T}^{0}}{c} = - \tau m \int f_{0} \vec{\Lambda} d^{2} p .
\end{equation}

Since the surface particle density is given by $\rho = \int f_{0} d^{2} p$, we have
\begin{equation}\label{appendixCurrent9}
\frac{\vec{T}^{0}}{c} = - \tau \rho m \vec{\Lambda} .
\end{equation} 

The final expression for the mass density current is
\begin{equation}\label{appendixCurrent10}
\vec{J}_{g} = - \sigma \vec{\Lambda},
\end{equation}
with mass conductivity coefficient for the mass $m$,
\begin{equation}\label{appendixCurrent11}
\sigma = \tau \rho m
\end{equation}
which completes the derivation.

\subsection{}
\label{appendixHalone}

Derivation of the equation for the gravitomagnetic field $\vec{H}_{g}$ eq. (\ref{Halone1}). 
For simplicity we introduce the notation $\alpha = \frac{16\pi G}{c^2}$. 
From Ohm's law for gravitoelectromagnetic fields,
eq. (\ref{lorentz0}), using the rotation operator and with help of eq. (\ref{gravmaxwell1b}), we have 

\begin{equation}\label{appendixHalone1}
-\partial_{t} \vec{H}_{g} = -\frac{\mu}{\sigma} \left(  \vec{\nabla} \times  \left( \vec{J}_{g} \times \vec{H}_{g} \right) \right) +
\frac{1}{\sigma} \left( \vec{\nabla} \times \vec{J}_{g} \right) .
\end{equation}
 
The part $\vec{\nabla} \times \vec{J}_{g}$ can be calculated from eq. (\ref{gravmaxwell1d}). Starting with   
\begin{equation}\label{appendixHalone2}
\vec{J}_{g} = -\frac{h}{\alpha} \left( \vec{\nabla} \times \vec{H}_{g} \right) 
\end{equation}
and taking the rotation operation on both sides we have
\begin{equation}\label{appendixHalone3}
\vec{\nabla} \times \vec{J}_{g} = -\frac{h}{\alpha} \left(  \vec{\nabla} \times  \left( \vec{\nabla} \times \vec{H}_{g} \right) \right)  .
\end{equation}
Substituting eq. (\ref{appendixHalone3}) into eq. (\ref{appendixHalone1}) we have

\begin{equation}\label{appendixHalone4}
-\partial_{t} \vec{H}_{g} = -\frac{\mu}{\sigma} \left(  \vec{\nabla} \times  \left( \vec{J}_{g} \times \vec{H}_{g} \right) \right)   
- \frac{h}{\alpha \sigma } \left(  \vec{\nabla} \times  \left( \vec{\nabla} \times \vec{H}_{g} \right) \right) .
\end{equation}
Using $\frac{\mu}{\sigma} = \frac{2}{m \rho}$, 
$\vec{\nabla}\times\left(\vec{\nabla}\times\vec{H}_{g}\right)=\vec{\nabla}\cdot\left(\vec{\nabla}\cdot\vec{H}_{g}\right)-\vec{\nabla}^2 \vec{H}_{g}$
and eq. (\ref{gravmaxwell1c}) ($\vec{\nabla} \vec{H}_{g} = 0$), we have the final form  

\begin{equation}\label{appendixHalone7}
\partial_{t} \vec{H}_{g} - \frac{2}{m \rho} \left(  \vec{\nabla} \times  \left( \vec{J}_{g} \times \vec{H}_{g} \right) \right) =
- \frac{h}{\alpha \sigma } \left( \vec{\nabla}^2 \vec{H}_{g} \right),
\end{equation}
which completes the derivation.


\newpage

\section{FIGURES}

\end{document}